\documentclass[prl,reprint,longbibliography]{revtex4-1}

\usepackage{amsmath}    % need for subequations
\usepackage{graphicx}   % need for figures
\usepackage{verbatim}   % useful for program listings
\usepackage{xcolor}      % use if color is used in text
\usepackage{subfigure}  % use for side-by-side figures
\usepackage{hyperref}   % use for hypertext links, including those to external documents and URLs

\usepackage[ugly]{nicefrac}

\usepackage{xfrac}

\usepackage{siunitx}

\usepackage{amssymb}
\usepackage{dsfont}

\usepackage{tabularx} % for appendix tables

\usepackage{pifont}

\usepackage{enumitem} % used for enumeration

\usepackage{notes2bib}

\usepackage{marginnote}

\bibnotesetup{
	note-name = ,
	use-sort-key = false
}

\definecolor{Grey}{RGB}{100,100,100}
\definecolor{QBlue}{RGB}{0,128,128}
\definecolor{LPurple}{RGB}{188,0,128}
\definecolor{ARed}{RGB}{255,85,85}
\definecolor{IYellow}{RGB}{255,200,50}

\def\flag{{\includegraphics[scale=0.35]{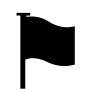}}}
\def\skull{{\includegraphics[scale=0.35]{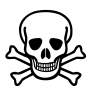}}}

\begin{document}	
	
\title{A reset-if-leaked procedure for encoded spin qubits}
\author{Veit Langrock}
\author{David P. DiVincenzo}
\affiliation{
Institute  for  Quantum  Information,\\ RWTH Aachen University,  D-52056 Aachen,  Germany\\ Peter Gr\"unberg Institute,  Theoretical Nanoelectronics, D-52425 J\"ulich,  Germany and \\J\"ulich-Aachen  Research  Alliance  (JARA), Fundamentals of Future Information Technologies, D-52425 J\"ulich, Germany
}
	
\begin{abstract}
    We report a substantially simplified procedure, based on group-theoretic reasoning, for the reduction of qubit leakage in exchange-only spin qubits. We to find exchange sequences which accomplish leakage reduction with only two additional spins and with as few as 14 nearest neighbor exchange interactions, less than half than previously reported. We show that the identified sequences are robust in the presence of realistic noise levels in the semiconductor environment. Our procedure also produces flag information that can be helpful in the implementation of quantum algorithms. 
\end{abstract}
	
\maketitle

\textit{Introduction} - While solid state qubits have made significant progress towards achieving large-scale quantum computation \cite{Arute2019}, an ever-present need has been to further increase the precision of quantum gate operations. Good progress continues to be made in the reduction of qubit noise rates (decoherence, relaxation), but a special concern is the process of leakage \cite{Plenio1997}, in which the system leaves the computational space. It is recognized that the introduction of leakage reduction units (LRUs) \cite{Preskill1998} must necessarily supplement the procedures of quantum error correction (QEC) for scalable quantum computation to be feasible. LRUs must be tailored to the specifics of the leakage process, and have, to varying degrees of specificity, been introduced for all major qubit types.

This letter reports a major improvement in the design of LRUs for gate-controlled spin qubits, with a twofold reduction in complexity of the LRU compared with previous ones for the exchange-only (EO) qubit. The EO qubit is promising because it permits extreme simplicity of qubit control: only one type of physical action, the controlled pulsing of the exchange interaction between electrons in neighboring quantum dots \cite{LossDiVincenzo1998,DiVincenzo2000}, can achieve universal quantum computation. The EO qubit is embodied in the states of three elementary spins. Its essential properties are determined by group theory; its data-carrying states correspond to the $j=\text{\sfrac{1}{2}}$ irreducible representations of the three-spin space. But its $j=\text{\sfrac{3}{2}}$ states are noncomputational, and inadvertent occupation of these states must be ameliorated by an LRU. While energy relaxation provides a natural mechanism to restore superconducting transmon qubits to a computational state over time \cite{Varbanov2020}, such processes tend to rather induce leakage in EO qubits due to the almost degeneracy of the $j=\text{\sfrac{1}{2}}$ and $j=\text{\sfrac{3}{2}}$ states in the characteristic EO operational regime \cite{Andrews2019}.

We call our LRU a ``Reset if Leaked” procedure (RiL).  Its innovation, compared with previous LRU strategies considered for spin qubits (repeated-CNOT LRU \cite{Preskill1998}, SWAP-if-Leaked LRU \cite{DiVincenzo2000,FongWandzura2011}), is that a very economical ancilla is needed – not even one qubit.  For the EO qubit, this means not even three electrons.  We find, using group-theoretic considerations, that exactly two electrons are sufficient to successfully implement RiL.  We numerically find exchange-pulse sequences that implement this RiL, with the number of pulses being half as many as needed with the full-qubit, SWAP-if-Leaked strategy.  This procedure is short enough that, according to the simulations we report here, it can be successful at currently available fidelity levels for silicon spin qubits \cite{Andrews2019}.      

\textit{Resources and Procedure} - 

For the EO qubit, leakage means that the system leaves the spin-\sfrac{1}{2} manifold, so that its state necessarily becomes one with angular momentum \sfrac{3}{2}$\hbar$.  Removal of leakage therefore requires the reduction of the total angular momentum of the three quantum dots by one unit of $\hbar$, necessarily via interaction with ancillary spins when constrained to exchange operations. Taking up a full unit of angular momentum requires the ancilla to have at least two spin-\sfrac{1}{2} particles.  We will show below that two single-electron quantum dots are indeed sufficient to reset an EO qubit to its computational space, improving on the three-dot ancilla used in previous \cite{FongWandzura2011,Kempe2001} leakage-reduction constructions.

\begin{figure}[h!]
	\centering
	\includegraphics[scale=0.3]{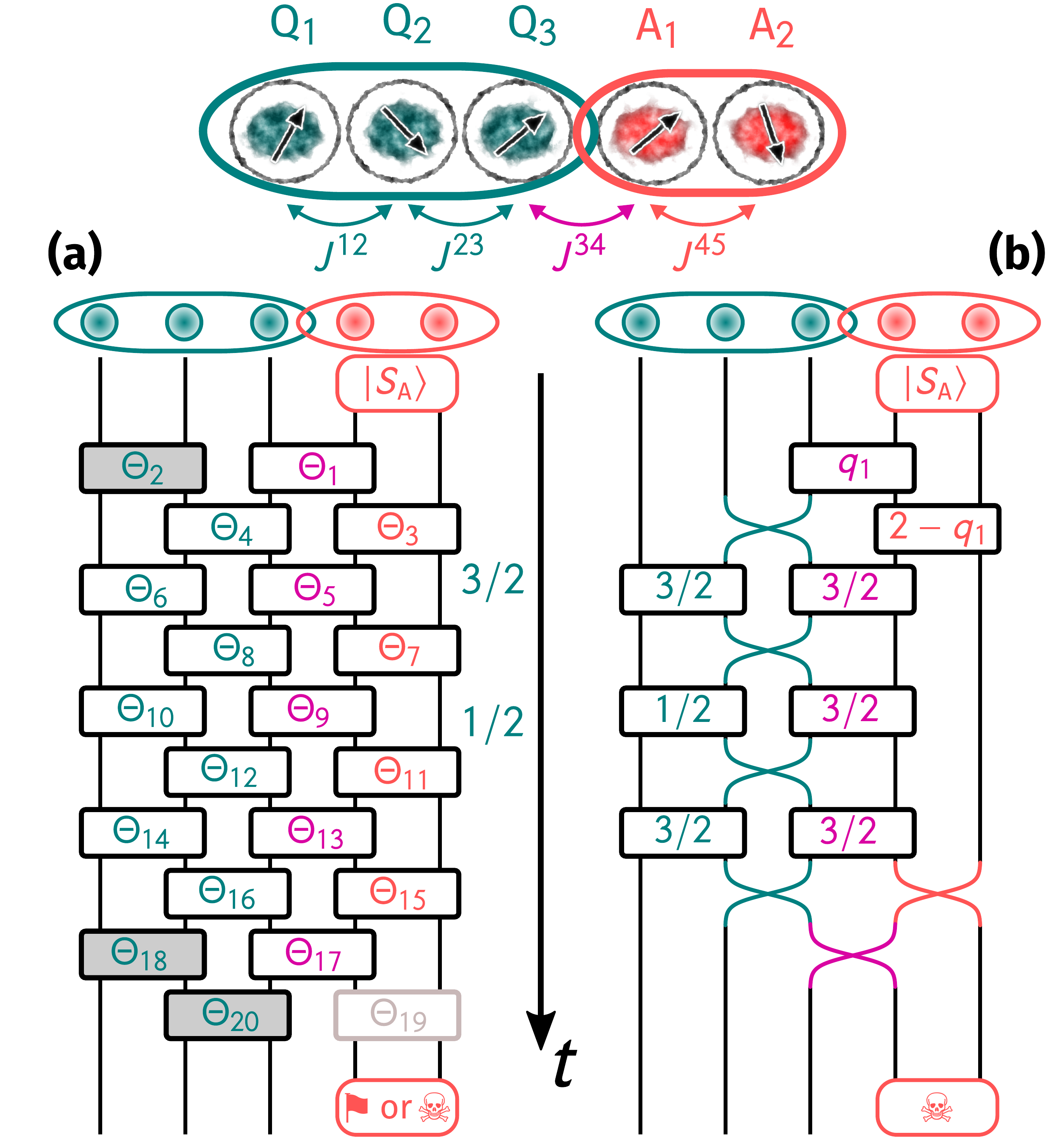}
	\caption{\label{fig:DotsAndSequence} Sketch of the five dot device with the corresponding exchanges. The generic brickwork exchange sequence with angles $\Theta$ ($\Theta^{ij}\equiv \int \mathrm{d}tJ^{ij}(t)$) is shown in \textbf{(a)}, with the gates used for realizing action \ref{numopt:5_makeQubitGateNice} indicated via grey background. Measurement and unconditional termination of the ancillary state are symbolized by 
	\protect\flag $\,$ and \protect\skull, respectively.
	The sequence in \textbf{(b)} is a solution lacking flaggability and with reset state $\cos(\pi/6)|{0}_\mathrm{Q}\rangle+\sin(\pi/6)|{1}_\mathrm{Q}\rangle$. The angle $q_1\pi=\arccos(1/3)$.}
\end{figure}

We will now (using the notation of \cite{Viola2001}), in stages, simplify the spin-representation structure  for the five spins (3 for qubit, 2 ancilla): First,
\begin{align}
&\big[\mathcal{D}_{1/2}^\text{Q1}\otimes\mathcal{D}_{1/2}^\text{Q2}\otimes\mathcal{D}_{1/2}^\text{Q3}\big]\otimes\big[\mathcal{D}^\text{A1}_{1/2}\otimes\mathcal{D}^\text{A2}_{1/2}\big]\nonumber\\
=&\big[\big(\mathcal{H}\otimes\mathcal{D}_{1/2}\big)^\text{Q}\oplus\mathcal{D}_{3/2}^\text{Q}\big]\otimes\big[\mathcal{D}^\text{A1}_{1/2}\otimes\mathcal{D}^\text{A2}_{1/2}\big]\label{eq:Product_Q_A}.
\end{align}
From here on, we label spin representations realized with the spins in the three qubit dots with superscript "Q" and spin realizations in the two ancillary dots with superscript "A", as indicated in Fig. \ref{fig:DotsAndSequence}.
The left three dots form the EO qubit, while the two dots on the right will serve as a disposal unit of leakage angular momentum. We now involve the two ancilla spin-\sfrac{1}{2} representations in the construction of the global spin spaces:
\begin{align}
=&\big\{\mathcal{H}^\text{Q}\otimes\big(\mathcal{D}_{1/2}^\text{Q}\otimes\mathcal{D}_{1/2}^\text{A1}\otimes\mathcal{D}_{1/2}^\text{A2}\big)\big\}\nonumber\\&\oplus\big(\mathcal{D}^\text{Q}_{3/2}\otimes\mathcal{D}^\text{A}_{0}\oplus\mathcal{D}^\text{Q}_{3/2}\otimes\mathcal{D}^\text{A}_{1}\big).\label{eq:Sum_12_32}
\end{align}
Here we have converted the direct product of the ancilla spins into their irreducible components ($\mathcal{D}^\text{A1}_{1/2}\otimes\mathcal{D}^\text{A2}_{1/2} = \mathcal{D}^\text{A}_{0}\oplus\mathcal{D}^\text{A}_{1}$) in the lower line. In the upper line, the total spin-\sfrac{1}{2} representation $\mathcal{D}_{1/2}^\text{Q}$ formed by the three computational dots provides a third spin-\sfrac{1}{2} particle on top of the two ancilla ones, which we can use to define another subsystem (EO) qubit:
\begin{align}
=&\big\{\mathcal{H}^\text{Q}\otimes\big[\big(\mathcal{H}\otimes\mathcal{D}_{1/2}\big)^\text{QA}\oplus\mathcal{D}_{3/2}^\text{QA}\big]\big\}\nonumber\\&\oplus\big(\mathcal{D}^\text{Q}_{3/2}\otimes\mathcal{D}^\text{A}_{0}\oplus\mathcal{D}^\text{Q}_{3/2}\otimes\mathcal{D}^\text{A}_{1}\big).
\end{align}
Rearranging and labeling the subspaces with subscripts corresponding to their total angular momentum $J$, we end up with:
\begin{align}
=&\overset{\text{\large\ding{182}}}{\left(\mathcal{H}^\text{Q}\otimes\mathcal{H}^\text{QA}\otimes\mathcal{D}^\text{QA}_{1/2}\right)_{\frac{1}{2}}}\nonumber\\&\oplus\overset{\text{\large\ding{183}}}{\big(\mathcal{D}^\text{Q}_{3/2}\otimes\mathcal{D}^\text{A}_{0}\big)_{\frac{3}{2}}}\oplus\overset{\text{\large\ding{184}}}{\big(\mathcal{H}^\text{Q}\otimes\mathcal{D}_{3/2}^\text{QA}\big)_{\frac{3}{2}}}\nonumber\\&\oplus\overset{\text{\large\ding{185}}}{\big(\mathcal{D}^\text{Q}_{3/2}\otimes\mathcal{D}^\text{A}_{1}\big)_{\frac{1}{2},\frac{3}{2},\frac{5}{2}}}\label{eq:Direct_Sum_structure_space}
\end{align}
which will serve as the basis for the following discussion. 
As guided by Eq. (\ref{eq:Direct_Sum_structure_space}), leakage reduction is achieved by designing an exchange sequence performing the following operations:
\begin{enumerate}[label=(\roman*)]	\item Prepare a singlet in the ancillary spins (i.e. initialize into $\mathcal{D}^\text{A}_{0}$) \label{num:1_ancillaInitSinglet}
	\item Don't entangle states in $\mathcal{H}^\text{Q}$ with $\mathcal{H}^\text{QA}$ in {\large\ding{182}} \label{num:2_donotEntangle}
	\item Transfer states from {\large\ding{183}} to {\large\ding{184}} \label{num:3_resetLeakedState}
	\item Don't end up in {\large\ding{185}} \label{num:4_donotInduceLeakage}
\end{enumerate}
The following specifications for the bipartite evolution enforced by \ref{num:2_donotEntangle} are optional:
\begin{enumerate}[label=(\roman*)]
	\setcounter{enumi}{4}
	\item In  {\large\ding{182}}, restrict the single qubit operation on $\mathcal{H}^\text{Q}$ to be a Pauli- or Clifford-type gate \label{numopt:5_makeQubitGateNice}
	\item In  {\large\ding{182}}, restrict the gate performed on $\mathcal{H}^\text{QA}$ so that the state initialized in \ref{num:1_ancillaInitSinglet} is mapped to itself \label{numopt:6_makeSequenceFlaggable}
	
\end{enumerate}
Action (v) reflects the thought that gates should form self-contained building blocks which may be arranged in a quantum circuit --- any overly complicated single qubit gate would require correction anyway.
Action (vi) accommodates the possibility of flagging a leaked qubit via singlet-triplet measurement of the ancillary states \cite{Bennett1997}, which we will discuss in some more detail below.
	
The sequence of actions described above is achieved via a specific succession of pulsed Heisenberg exchanges between the individual dots ($\hat{H}^{ij}(t)=J^{ij}(t)\boldsymbol{\hat{S}}_i\cdot\boldsymbol{\hat{S}}_j$), laid out in a linear nearest-neighbor coupled array with pairwise interactions arranged in a "brickwork" space-time layout (cf. \textbf{Fig. \ref{fig:DotsAndSequence} (a)} and \textbf{Appendices \ref{sec:AppStates}} and \textbf{\ref{sec:AppExchangeGates}}), as done in past work \cite{DiVincenzo2000, FongWandzura2011, Zeuch2016,Andrews2019}. The pulse sequences are found via a numerical optimization routine, the details of which are referred to \textbf{Appendix \ref{sec:AppNumercialSearch}}.

\textit{Resulting sequences} - The shortest sequence found, performing actions \ref{num:1_ancillaInitSinglet}-\ref{num:4_donotInduceLeakage}, is displayed in \textbf{Fig. \ref{fig:DotsAndSequence} (b)}, using 14 exchange gates over 9 brickwork layers -- about half the length of the previously found leakage reduction circuit \cite{FongWandzura2011}. Our sequence also completes action \ref{numopt:5_makeQubitGateNice} by performing an identity operation on an unleaked computational qubit. But the sequence fails to implement action \ref{numopt:6_makeSequenceFlaggable}, since the output in the total spin-1 subspace in the ancillas is a triplet (i.e. $\mathcal{D}^\text{A}_{1}$).
	
A sequence capable of flagging leakage requires two more exchanges to achieve \ref{numopt:6_makeSequenceFlaggable} and three more exchanges and one more layer to fulfill \ref{numopt:5_makeQubitGateNice}, which may again be chosen as identity. With 19 exchanges over 10 brickwork layers, this uses all exchanges except $\Theta_{19}$ displayed in \textbf{Fig. \ref{fig:DotsAndSequence} (a)}.
	
Even with the constraints of minimal number of gates, the chosen positions of the gates, full flaggability, and identity on the computational qubit, we determine there to be no fewer than 264 distinct solutions. These differ only in the reset state to which the qubit is reinitialized after leakage. The details of the sequences and a select set of solutions are presented in \textbf{Appendix \ref{sec:AppSolutions}}.
	
\textit{Noise characterization} - To compare the different implementations and discriminate between the multitude of equivalent solutions in the flag sequence case, we study sequences where the exchange couplings are subject to noise, yielding slightly imperfect realizations.
	
We assume that the dominant noise mechanism affecting the fidelity of the studied sequences are of charge noise type, which translates into noise of the exchange couplings between nearby quantum dots. We do not consider effective magnetic noise leading to different spin precession frequencies between dots, as measures such as isotopic purification pave clear pathways to reducing such noise influence in the future.

We therefore study a noisy channel that can be expressed as an ensemble average of slightly imperfect realizations of the gate sequence:
\begin{equation}
	\mathcal{E}(\rho) = \int \mathrm{d}\boldsymbol{x} \, p(\boldsymbol{x})\,\hat{T}[\boldsymbol{\Theta}(\boldsymbol{x})]\rho \hat{T}^\dagger[\boldsymbol{\Theta}(\boldsymbol{x})].
	\label{eq:noisyChannelGeneral}
\end{equation}	
Here, $\hat{T}(\boldsymbol{\Theta})$ is the isometry realized by the exchange sequence $\boldsymbol{\Theta}$, with each individual $\Theta_k\equiv \int \mathrm{d}t J^{ij}(t)$ being the exchange angle realized by the exchange pulse $J^{ij}(t)$ (refer to \textbf{Appendix \ref{sec:AppExchangeGates}} for details, including Eq. (\ref{seqq}) for the numbering convention) and  $\boldsymbol{x}\in \mathbb{R}^{N}$ is a random vector representing the different noise realizations ($N=20$ is the maximum number of employed exchange gates).
	
We assume that charge noise expresses itself as a shift in amplitude of the exchange coupling, and that the temporal shape of the pulse remains mostly undistorted. The resulting elements of the exchange vector $\boldsymbol{\Theta}(\boldsymbol{x})$ are then given as $\Theta_k(\boldsymbol{x}) = \Theta_{0,k}\times(1+x_k)$ in our model, with $\boldsymbol{\Theta}_0$ the unperturbed exchange sequence, making $\hat{T}(\boldsymbol{\Theta}_0)$ the isometry of the ideal channel. The locations of the exchanges correspond to the gate layout indicated in Fig. \ref{fig:DotsAndSequence} (left).
In this study, we assume fully non-Markovian noise behavior, i.e. exchange noise remaining static over the whole duration of the sequence, motivated by the comparatively short duration of such exchange sequences \cite{Andrews2019} and by charge noise usually showing very non-Markovian behavior due to its $1/f$-character, i.e. dominant noise power spectral weight at low frequencies \cite{Paladino2014}.
Further assuming all exchange noise to be spatially uncorrelated, we may then write the joint probability distribution in (\ref{eq:noisyChannelGeneral}) as 
\begin{align}
	p(\boldsymbol{x})=\prod_{i=1}^4 \left[ p_\mathrm{g}(x_i)\prod_{n=1}^{N/4-1}\delta(x_i-x_{i+4n})\right]
	\label{eq:fullNonMarkovModel}
\end{align}	
$p_\mathrm{g}(x)$ is modeled as a normal distribution of zero mean and standard deviation $\sigma$ \bibnote{Technically $p_\mathrm{g}(x)$ should be folded around $-1$ such that $\boldsymbol{\Theta}(\boldsymbol{x})$ is strictly positive, but noise strengths where this modelling detail becomes relevant should not be representative of real devices.}, the latter being a measure of the characteristic charge noise strength.
Further details can be found in \textbf{Appendices \ref{sec:AppSolutions} and \ref{sec:AppNoiseStudyDetails}}.

To aid the following discussion, we fix states in the Hilbert space structure in {\large\ding{182}}. The computational qubit states are defined by the representation in the three Q dots $\mathcal{D}^\text{Q}_{1/2}$, i.e. $\big(\mathcal{D}_{1/2}^\text{Q1}\otimes\mathcal{D}_{0}^\text{Q23}\big)_{\frac{1}{2}}\equiv |{0}_\mathrm{Q}\rangle$ and $\big(\mathcal{D}_{1/2}^\text{Q1}\otimes\mathcal{D}_{1}^\text{Q23}\big)_{\frac{1}{2}}\equiv |{1}_\mathrm{Q}\rangle$. The auxillary qubit states are determined by the states in the two ancillary dots, $\big(\mathcal{D}^\text{Q}_{1/2}\otimes\mathcal{D}^\text{A}_{0}\big)_{\frac{1}{2}}\equiv|{0}_\mathrm{QA}\rangle$ and $\big(\mathcal{D}^\text{Q}_{1/2}\otimes\mathcal{D}^\text{A}_{1}\big)_{\frac{1}{2}}\equiv|{1}_\mathrm{QA}\rangle$ \bibnote{The symmetry in structure between the Q and QA qubits is no coincidence, as the QA qubit can be viewed as yet another EO qubit using the total spin (gauge) of the (unleaked) Q qubit as a third spin in addition to the ancillary dots. From this perspective, we still perform a SWAP-if-leaked operation, as the QA qubit ends up in the usual EO leakage states after a successful RiL sequence.}. A detailed listing of all states and their explicit spin configurations may be found in \textbf{Appendix \ref{sec:AppStates}}, and further details on the Q and AQ subsystem structure and the operation on it can be found in \textbf{Appendix \ref{sec:AppNumercialSearch}}.
	A fiducial RiL sequence starts with its input in the state $|{0}_\mathrm{QA}\rangle$ and explores only the four-dimensional state space in {\large\ding{182}}, while fulfilling action \ref{num:2_donotEntangle}.
	Faulty operation may either lead to a different outcome in $\mathcal{H}_\mathrm{QA}$, making the task of flagging leakage more difficult, or may change the representation of $\mathcal{D}^\text{Q}_{1/2}$, which corresponds to a qubit error, or may even transfer us to the representation $\big(\mathcal{D}^\text{Q}_{3/2}\otimes\mathcal{D}^\text{A}_{1}\big)_\frac{1}{2}$ in {\large\ding{185}}, which corresponds to inducing leakage in the computational qubit.
	To characterize error rates, we focus on the spin-$\sfrac{1}{2}$ sector, as we assume that any qubit worth saving has significantly more support in the computational states than in the leaked ones.
	For this purpose, we further introduce projectors onto the qubit state being leaked or unleaked, $P_\mathrm{Q}$ on $\mathcal{D}^\text{Q}_{1/2}$ and $P_\mathrm{L}$ on $\mathcal{D}^\text{Q}_{3/2}$, $P_\mathrm{Q}+P_\mathrm{L}=I$. These divide the Hilbert space into qubit and leakage subspaces $\mathcal{H}_\mathrm{Q}\oplus\mathcal{H}_\mathrm{L}$ of the computational EO qubit. When applied after the projector onto the global spin-$\sfrac{1}{2}$ ($J=\sfrac{1}{2}$) space $P_{\sfrac{1}{2}}$, $P_\mathrm{Q}$ and $P_\mathrm{L}$ project onto {\large\ding{182}} and the spin-$\sfrac{1}{2}$ component of {\large\ding{185}}, respectively. 
	Performing finally the trace over the ancillary spin states (see \textbf{Appendix \ref{sec:AppCohLeakage}} for details), this yields the qubit channel of interest:
\begin{align}
\widetilde{\mathcal{E}}(\rho_\mathrm{Q}) &= \mathrm{Tr}_\mathrm{A}\left[\mathcal{E}(\rho_\mathrm{Q}\oplus 0_\mathrm{L})\right],
\end{align}
with $\rho_\mathrm{Q}$ constrained to unleaked qubit density matrices ($\widetilde{\mathcal{E}}:\;\mathcal{H}_\mathrm{Q}\rightarrow \mathcal{H}_\mathrm{Q}\oplus\mathcal{H}_\mathrm{L}$), implying that we only consider evolution in the global $J=\sfrac{1}{2}$ subspace.
	
\begin{figure}%[h!]
	\includegraphics[scale=0.6]{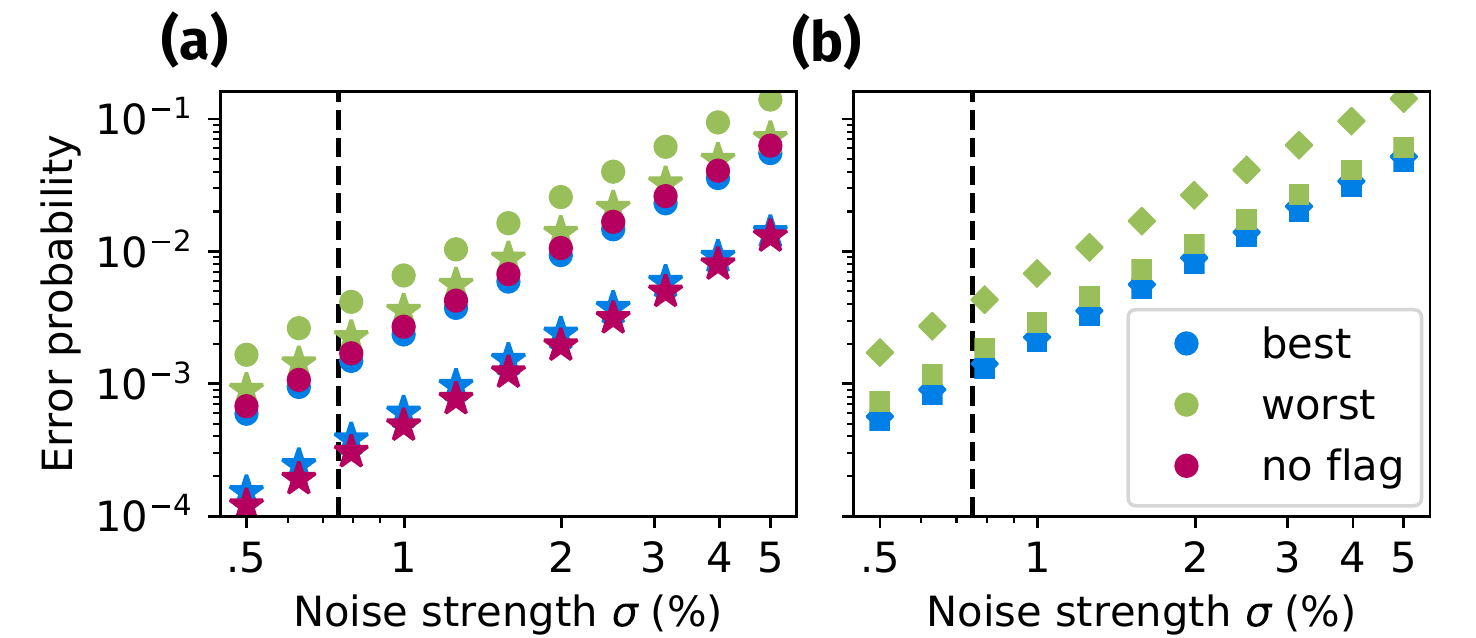}
	\caption{\label{fig:noisyChannelPlots} Plots comparing average induced leakage $\bar{p}_\mathrm{L,ind}$ ({\ding{72}}) and gate infidelity $1-\bar{F}_\mathrm{Q}$ ({\scriptsize\ding{108}}) for different realizations of the RiL sequence \textbf{(a)} and the average flag error $\bar{\epsilon}_\mathrm{F}$ ({\ding{117}}) and failure to transfer a leaked state from  {\large\ding{183}} $\bar{\epsilon}_5$ ({\scriptsize\ding{110}}) for the flaggable solutions \textbf{(b)} with respect to varying noise strength $\sigma$. The dashed line marks a noise strength of $\sigma=1/(\sqrt{2}\pi N_\mathrm{coh})\approx0.75\%$ matched to the number of $N_\mathrm{coh}>30$ coherent exchange oscillations reported in \cite{Reed2016}.}
\end{figure}

Since the primary role of the proposed procedure is the removal of leakage, an important figure of merit is how much leakage, on average, is induced in an unleaked qubit by execution of the sequence under noisy operation, counteracting its intended use.
Application of a leakage removal procedure only makes sense if the input accumulated leakage $\epsilon_\mathrm{L}$ is appreciably higher than leakage induced by the sequence, i.e. $\epsilon_\mathrm{L}\gg \bar{p}_\mathrm{L,ind}$, where
\begin{equation}
\bar{p}_\mathrm{L,ind}=\mathrm{Tr}\big[P_\mathrm{L}\widetilde{\mathcal{E}}\left(I_\mathrm{Q}/2\right)\big].\label{firsterror}
\end{equation}
Here $I_\mathrm{Q}/2$ is the maximally mixed state in the unleaked qubit subsystem. 
For effective leakage application, the induced leakage rates in \textbf{Fig. \ref{fig:noisyChannelPlots} (a)} denote a practical lower bound on the input leakage probability, and therefore allow an estimate for the frequency with which such RiL procedures should be carried out to be effective, ideally based on experimental leakage rate benchmarks such as acquired in \cite{Andrews2019}.

	Even if the qubit stays in the computational subspace, qubit errors may still occur. We characterize this by computing the average gate fidelity for the qubit channel with leakage $\widetilde{\mathcal{E}}(\cdot)$ (non-trace preserving in the qubit subspace), which is given by \cite{WoodGambetta2018}
	\begin{equation}
	 \bar{F}_\mathrm{Q} =\frac{d_\mathrm{Q}F_\mathrm{e}+1-\bar{p}_\mathrm{L,ind}}{d_\mathrm{Q}+1},\label{seconderror}
	\end{equation}
	with the entanglement fidelity $F_\mathrm{e}$ and the dimension of the qubit subspace $d_\mathrm{Q}=2$
	\bibnote{The entanglement fidelity may be compactly computed for the qubit channel $\widetilde{\mathcal{E}}(\cdot)$ via some associated set of Kraus operators $\{\widetilde{E}_i\}$, evaluated for a maximally mixed state in the qubit subspace $\rho_\mathrm{Q} = I_\mathrm{Q}/d_\mathrm{Q}$, yielding \cite{WoodGambetta2018,Schumacher1996,NielsenChuang2010}:
	\begin{equation}
	F_\mathrm{e} = \frac{1}{d_\mathrm{Q}^2}\sum_i|\mathrm{Tr}(\mathcal{I}^\dagger\widetilde{E}_iI_\mathrm{Q})|^2,
	\end{equation}
	with isometry $\mathcal{I}=I_\mathrm{Q}\oplus 0_\mathrm{L} $ the identity on the qubit subspace.}.

	We may write the average qubit fidelity as $\bar{F}_\mathrm{Q}=(1-\bar{p}_\mathrm{L,ind})\bar{F}_2$ with $\bar{F}_2$ the average fidelity of operation in the computational subspace. For small errors, $1-\bar{F}_2\approx(1- \bar{F}_\mathrm{Q})-\bar{p}_\mathrm{L,ind}$ is a measure of how much gate errors in the qubit subspace contribute to the total infidelity, while $\bar{p}_\mathrm{L,ind}$ is the contribution due to leakage. Qubit errors in the computational subspace may be corrected afterwards by means of canonical quantum error correction, and are therefore seen as a less hazardous contribution to the qubit infidelity than $\bar{p}_\mathrm{L,ind}$. \textbf{Fig. \ref{fig:noisyChannelPlots} (a)}  displays these quantities against varying noise strengths of expected experimental relevance (error rates for a leaked input are given in \textbf{Appendix \ref{sec:AppSolutions}}, while the arguably more interesting aspect of error coherence is referred to \textbf{Appendix \ref{sec:AppCohLeakage}}). One may observe that the least error prone flaggable solution, despite the longer gate sequence, displays a negligible difference in erroneous operation compared to the unflaggable solution displayed in \textbf{Fig. \ref{fig:DotsAndSequence} (a)}.
	We will therefore now discuss further the implications of being able to flag leakage in different settings.

\textit{Flaggability} - Identifying if a qubit suffered leakage or not, even when leakage is removed, provides useful information in both the near term setting of Noisy Intermediate Scale Quantum (NISQ) applications as well as in future Quantum Error Corrected (QEC) processors.
Associating the flag result $0_\mathrm{M}$ with an \textit{unleaked} input that is transferred to an \textit{unleaked} output ($\mathrm{U}_\mathrm{out}\mathrm{U}_\mathrm{in}$) and result $1_\mathrm{M}$ with a \textit{leaked} input which is reset to an \textit{unleaked} output ($\mathrm{U}_\mathrm{out}\mathrm{L}_\mathrm{in}$), the probabilites of making the wrong guess given a flag result are (see \textbf{Appendix \ref{sec:AppFlaggingConditionalProbabilites}} for details):
\begin{align}
P(\overline{U_\mathrm{out}U_\mathrm{in}}|0_\mathrm{M})&\approx \epsilon_{0T}\bar{\epsilon}_\mathrm{L,ind}+\epsilon_{0T}\epsilon_\mathrm{L}+\bar{\epsilon}_{5}\epsilon_\mathrm{L},
\label{eq:FalsePositive}\\
P(\overline{U_\mathrm{out}L_\mathrm{in}}|1_\mathrm{M}) &\approx \frac{1}{1+\epsilon_\mathrm{L}/(\epsilon_{1S}+\bar{\epsilon}_\mathrm{F})}.
\label{eq:FalseNegative}\\
\bar{\epsilon}_\mathrm{F} = \mathrm{Tr}&\left[\left( I-|0_\mathrm{QA}\rangle \langle 0_\mathrm{QA}|\right)P_{1/2}\mathcal{E}\left(P_\mathrm{Q}/d_\mathrm{Q}\right)\right]\label{thirderror}	\end{align}
is the average probability of failure to remain in the QA state $|{0}_\mathrm{QA}\rangle$ \bibnote{Note that we use the 5 spin channel $\mathcal{E}\left(\cdot\right)$ here, as the state of the ancillary spins is the quantity of interest}, plotted in Fig. \ref{fig:noisyChannelPlots} (right) for selected solutions. The conditional probabilities $\epsilon_{1S}\equiv P(1_\mathrm{M}|S_\mathrm{A})$ and $\epsilon_{0T}\equiv P(0_\mathrm{M}|T_\mathrm{A})$ are phenomenological measurement-error probabilities of misidentifying the ancilla state and
$\bar{\epsilon}_5$ is the probability that transfer of a leaked state from {\large\ding{183}} fails, so that the state of the Q spins remains leaked and the ancillas remain in the singlet state.
False positives (\ref{eq:FalsePositive}) should be unlikely, as they arise from joint occurence of both a leakage fault as well as an error in the sequence or measurement. If the noise causing faults is of Heisenberg exchange character as in (\ref{eq:noisyChannelGeneral}),
even leakage induced by the RiL sequence will usually lead to a positive flag result, as the $J=\text{\sfrac{1}{2}}$ state of {\large\ding{185}} is paired with a triplet ancilla state in this case, disallowing fautly RiL execution from introducing unflagged leakage into the qubit (even if the leakage flagged here is introduced after the fact). See \textbf{Appendix \ref{sec:AppFlaggingConditionalProbabilites}} for details.

In a NISQ setting, the burden of encoding faithful quantum information is put entirely on the physical EO qubit and a leakage event is a complete loss of qubit information. Flagging is therefore essential to allow for removal of leaked qubits from the computation, while a timely reset is not strictly necessary. For this purpose, leakage detection units would suffice, which may be constructed on a 5 spin-\sfrac{1}{2} basis in a manner analogous to the RiL sequences featured here \bibnote{A pure leakage detection procedure would consist of replacing the mapping {\large\ding{183}}$\rightarrow${\large\ding{184}} in \ref{num:3_resetLeakedState} by {\large\ding{183}}$\rightarrow${\large\ding{184}}$\oplus${\large\ding{185}} while leaving the ancillary qubit in state $|{0}_\mathrm{QA}\rangle$ in {\large\ding{182}} or, alternatively, mapping {\large\ding{183}} to itself and letting the ancillary qubit output be $|{1}_\mathrm{QA}\rangle$  in {\large\ding{182}}. This would be more along the lines of standard QEC ideas, allowing multiple shots at measurements since leakage is not actively removed. This might have applications for near term NISQ devices, but would not be practical for error corrected qubits, as active leakage correction is required to allow for operation of the QEC code, in contrast to qubit errors where knowledge about errors does not necessitate active manipulation. A search indicates that there is no significantly shorter sequence that performs this pure leakage detection.}.
In a QEC setting, the presence of leakage may already be inferred by decoding syndrome information \cite{Varbanov2020} given suitable two qubit interaction (see \textbf{Appendix \ref{sec:AppFlagQEC}} for a discussion of the QEC setting for EO qubits). LRU flags may provide a direct way of complementing this information, increasing the robustness of logical quantum information with respect to leakage faults, but are not strictly necessary. On the other hand, the presence of leaked qubits will degrade code performance \cite{Fowler2013}, making frequent removal of leakage a necessary operation fulfilled by the RiL sequence presented above.

In conclusion, we provide a physically motivated means of removing leakage in exchange only qubits, significantly reducing the overhead involved in their execution. We discuss and quantify error resilience as well as the capabilities of providing flag measurements.
It remains to be studied in detail how such gadgets would perform in envisioned many qubit applications. For example, in a NISQ setting, leakage detection may be performed at certain check points in a long running computation --- if many identical states need to be prepared for a subsequent computation, leakage detection sequences may help in post-selecting faulty qubit blocks and preventing them from spoiling the subsequent computation. For QEC, studies investigating error correction performance utilizing LRUs \cite{Suchara2015} under realistic parameters \cite{Rispler2020}, may prove insightful for the optimal utilization of RiL units. Finally, studying their performance under magnetic noise will provide further insight into expected experimental performance, as magnetic gradients still prove a significant source of error in current experimental setups \cite{Andrews2019}.

\textit{Acknowledgments} - We would like to thank Daniel Zeuch and for stimulating discussions. VL would like to thank Maximilian Russ for discussions on the current reality of spin qubits and related noise issues and members of the HRL group for insightful discussion at the 2019 Silicon Quantum Electronics Workshop.

We acknowledge support from the Deutsche Forschungsgemeinschaft (DFG, German Research Foundation) under Germany's Excellence Strategy Cluster of Excellence Matter and Light for Quantum Computing (ML4Q) EXC 2004/1 390534769.

\bibliography{GateSeq_Ref.bib}

\begin{thebibliography}{40}
\bibitem [{1()}]{1}%
  \BibitemOpen
  \href@noop {} {}\bibinfo {note} {Technically $p_\mathrm {g}(x)$ should be
  folded around $-1$ such that $\boldsymbol {\Theta }(\boldsymbol {x})$ is
  strictly positive, but noise strengths where this modelling detail becomes
  relevant should not be representative of real devices.}\BibitemShut {Stop}%
\bibitem [{2()}]{2}%
  \BibitemOpen
  \href@noop {} {}\bibinfo {note} {The symmetry in structure between the Q and
  QA qubits is no coincidence, as the QA qubit can be viewed as yet another EO
  qubit using the total spin (gauge) of the (unleaked) Q qubit as a third spin
  in addition to the ancillary dots. From this perspective, we still perform a
  SWAP-if-leaked operation, as the QA qubit ends up in the usual EO leakage
  states after a successful RiL sequence.}\BibitemShut {Stop}%
\bibitem [{4()}]{4}%
  \BibitemOpen
  \href@noop {} {}\bibinfo {note} {Note that we use the 5 spin channel
  $\mathcal {E}\left (\cdot \right )$ here, as the state of the ancillary spins
  is the quantity of interest}\BibitemShut {NoStop}%
\bibitem [{5()}]{5}%
  \BibitemOpen
  \href@noop {} {}\bibinfo {note} {A pure leakage detection procedure would
  consist of replacing the mapping {\large \ding {183}}$\rightarrow ${\large
  \ding {184}} in \ref {num:3_resetLeakedState} by {\large \ding
  {183}}$\rightarrow ${\large \ding {184}}$\oplus ${\large \ding {185}} while
  leaving the ancillary qubit in state $|{0}_\mathrm {QA}\rangle $ in {\large
  \ding {182}} or, alternatively, mapping {\large \ding {183}} to itself and
  letting the ancillary qubit output be $|{1}_\mathrm {QA}\rangle $ in {\large
  \ding {182}}. This would be more along the lines of standard QEC ideas,
  allowing multiple shots at measurements since leakage is not actively
  removed. This might have applications for near term NISQ devices, but would
  not be practical for error corrected qubits, as active leakage correction is
  required to allow for operation of the QEC code, in contrast to qubit errors
  where knowledge about errors does not necessitate active manipulation. A
  search indicates that there is no significantly shorter sequence that
  performs this pure leakage detection.}\BibitemShut {Stop}%
\end{thebibliography}

\clearpage
	
\appendix

\section{Explicit state description}
\label{sec:AppStates}

\begin{figure}[h!]
	%\centering
	\includegraphics[scale=0.3]{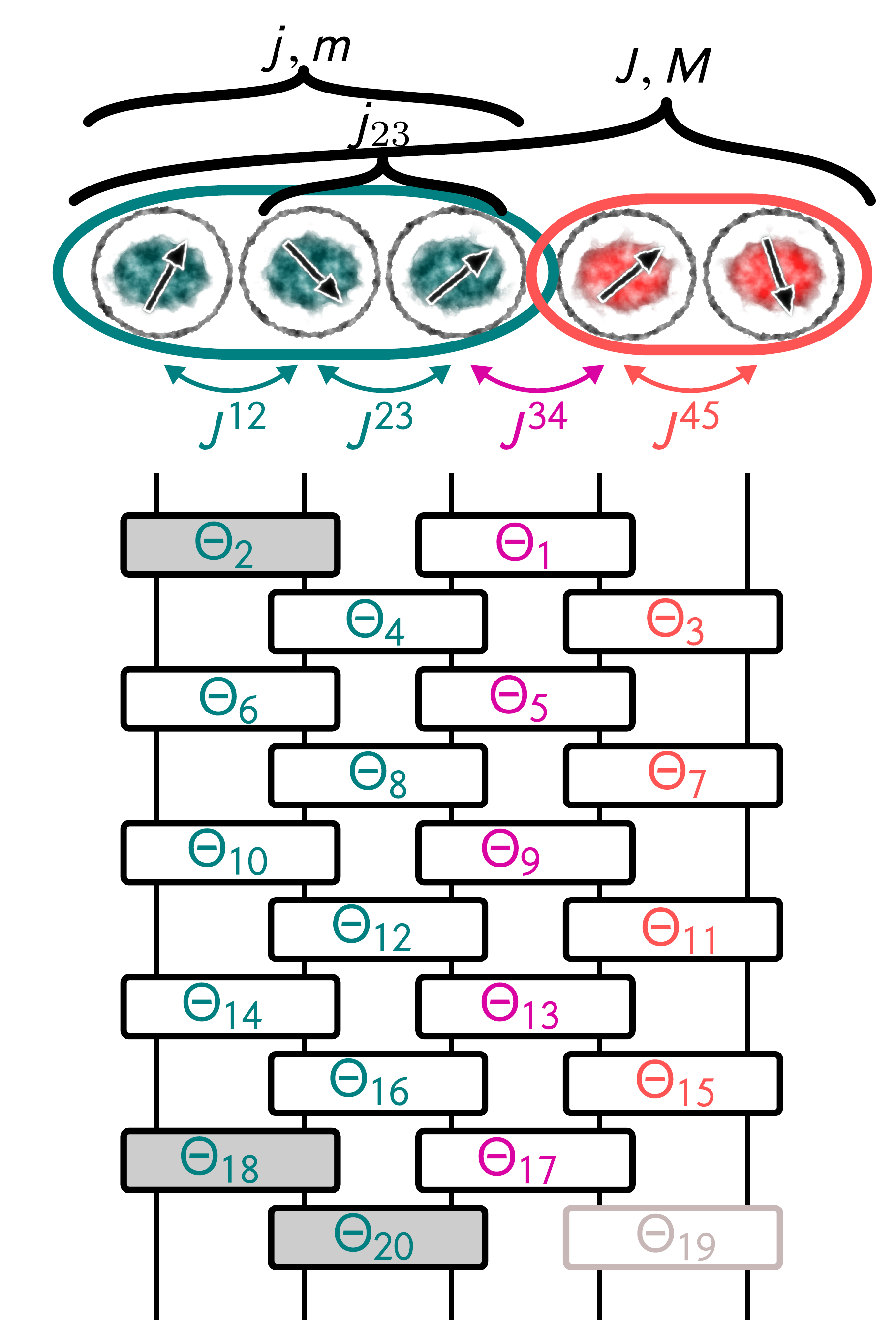}
	\caption{\label{fig:DotsBrickworkAppendix} Sketch of the employed exchanges and brickwork structure, reproduced from the main text as visual aid for state discussion and the solutions presented in \textbf{Tab. \ref{tab:AppExchangeTable}}.}
\end{figure}

In the following, we define a full set of states for the total spin spaces $J=1/2$ (see Fig. \ref{fig:DotsBrickworkAppendix} for the association between the quantum numbers used and the spin location)
\begin{alignat}{3}
\big(\mathcal{D}_{1/2}^\text{Q1}\otimes\mathcal{D}_{0}^\text{Q23}\big)_{\frac{1}{2}}&\otimes\mathcal{D}^\text{A}_{0}&&\equiv |0\rangle\label{A1} \\
\big(\mathcal{D}_{1/2}^\text{Q1}\otimes\mathcal{D}_{1}^\text{Q23}\big)_{\frac{1}{2}}&\otimes\mathcal{D}^\text{A}_{0}&&\equiv |1\rangle\\
\big(\mathcal{D}_{1/2}^\text{Q1}\otimes\mathcal{D}_{0}^\text{Q23}&\otimes\mathcal{D}^\text{A}_{1}\big)_{\frac{1}{2}}&&\equiv |2\rangle\\
\big[\big(\mathcal{D}_{1/2}^\text{Q1}\otimes\mathcal{D}_{1}^\text{Q23}\big)_{\frac{1}{2}}&\otimes\mathcal{D}^\text{A}_{1}\big]_{\frac{1}{2}}&&\equiv |3\rangle\\
\big(\mathcal{D}^\text{Q}_{3/2}&\otimes\mathcal{D}^\text{A}_{1}\big)_{\frac{1}{2}}&&\equiv |4\rangle
\end{alignat}
and for $J=3/2$
\begin{alignat}{3}
\mathcal{D}^\text{Q}_{3/2}&\otimes\mathcal{D}^\text{A}_{0}&&\equiv |5\rangle\\
\big(\mathcal{D}_{1/2}^\text{Q1}\otimes\mathcal{D}_{0}^\text{Q23}&\otimes\mathcal{D}^\text{A}_{1}\big)_{\frac{3}{2}}&&\equiv |6\rangle\\
\big[\big(\mathcal{D}_{1/2}^\text{Q1}\otimes\mathcal{D}_{1}^\text{Q23}\big)_{\frac{1}{2}}&\otimes\mathcal{D}^\text{A}_{1}\big]_{\frac{3}{2}}&&\equiv |7\rangle\\
\big(\mathcal{D}^\text{Q}_{3/2}&\otimes\mathcal{D}^\text{A}_{1}\big)_{\frac{3}{2}}&&\equiv |8\rangle.\label{A8}
\end{alignat}

Below we will write out the representations in the $J=1/2$ and $J=3/2$ sector for one specific spin projection quantum number $M$. But since the Heisenberg interaction commutes with the ladder operator $J_+=J_x+iJ_y$, this interaction will generate the evolution for the states specified above in any sector $M$. We have the 5 states with $z$-projection of angular momentum $M=-1/2$ as the representative states of the $J=1/2$ sector (the $M=+1/2$ states are generated by application of $J_+$) \cite{ClebschGordanTable}:
\begin{align}
\left|0_{-1/2}\right\rangle &= \left| \sfrac{1}{2},-\sfrac{1}{2},{0}_\text{Q}\right\rangle_\text{Q}|S\rangle_\text{A}\label{A10} \\
\left|1_{-1/2}\right\rangle &= \left|\sfrac{1}{2},-\sfrac{1}{2},{1}_\text{Q}\right\rangle_\text{Q}|S\rangle_\text{A}\label{A11} \\
\left|2_{-1/2}\right\rangle &= -\sqrt{\frac{1}{3}}\left|\sfrac{1}{2},-\sfrac{1}{2},{0}_\text{Q}\right\rangle_\text{Q}|T_0\rangle_\text{A}\nonumber\\&\;\;\;\;+\sqrt{\frac{2}{3}}\left|\sfrac{1}{2},+\sfrac{1}{2},{0}_\text{Q}\right\rangle_\text{Q}|T_-\rangle_\text{A} \label{A12}\\
\left|3_{-1/2}\right\rangle &= -\sqrt{\frac{1}{3}}\left|\sfrac{1}{2},-\sfrac{1}{2},{1}_\text{Q}\right\rangle_\text{Q}|T_0\rangle_\text{A}\nonumber\\&\;\;\;\;+\sqrt{\frac{2}{3}}\left|\sfrac{1}{2},+\sfrac{1}{2},{1}_\text{Q}\right\rangle_\text{Q}|T_-\rangle_\text{A} \label{A13}\\
\left|4_{-1/2}\right\rangle &= +\sqrt{\frac{1}{2}}\left|\sfrac{3}{2},-\sfrac{3}{2}\right\rangle_\text{Q}|T_+\rangle_\text{A}\nonumber\\&\;\;\;\;-\sqrt{\frac{1}{3}}\left|\sfrac{3}{2},-\sfrac{1}{2}\right\rangle_\text{Q}|T_0\rangle_\text{A}+\sqrt{\frac{1}{6}}\left|\sfrac{3}{2},+\sfrac{1}{2}\right\rangle_\text{Q}|T_-\rangle_\text{A}\label{A14}
\end{align}
and the 4 states with $M=-3/2$ as representative states of the $J=3/2$ sector (the $M=-1/2,\,1/2,\,3/2$ states are again generated by repeated application of $J_+$):
\begin{align}
\left|5_{-3/2}\right\rangle &= \left|\sfrac{3}{2},-\text{\sfrac{3}{2}}\right\rangle_\text{Q}|S\rangle_\text{A}\label{A15} \\
\left|6_{-3/2}\right\rangle &= \left|\sfrac{1}{2},-\text{\sfrac{1}{2}},{0}_\text{Q}\right\rangle_\text{Q}|T_-\rangle_\text{A} \label{A16}\\
\left|7_{-3/2}\right\rangle &= \left|\sfrac{1}{2},-\sfrac{1}{2},{1}_\text{Q}\right\rangle_\text{Q}|T_-\rangle_\text{A} \label{A17}\\
\left|8_{-3/2}\right\rangle &= -\sqrt{\frac{3}{5}}\left|\sfrac{3}{2},-\sfrac{3}{2}\right\rangle_\text{Q}|T_0\rangle_\text{A}+\sqrt{\frac{2}{5}}\left|\sfrac{3}{2},-\sfrac{1}{2}\right\rangle_\text{Q}|T_-\rangle_\text{A}
\end{align}

\section{Unitary exchange action}
\label{sec:AppExchangeGates}
To compute the local action of the Heisenberg interaction, we express it in terms of projectors on the singlet-triplet states of the dot pair in question:
$$U_{ij}(\Theta) = e^{i\Theta}|S_{ij}\rangle\langle S_{ij}|+|T_{ij}\rangle\langle T_{ij}|,$$
with $|T\rangle\langle T|$ being shorthand for the projector onto all triplet states $|T_{-}\rangle\langle T_{-}|+|T_{0}\rangle\langle T_{0}|+|T_{+}\rangle\langle T_{+}|$.
Without loss of generality, we consider the spin $J=1/2,M=-1/2$ and $J=3/2,M=-3/2$ blocks of the total spin system (again, the other $M$ states may be obtained by application of the ladder operator $J_+$).
Since the Heisenberg interaction conserves total spin angular momentum, we may decompose the global unitary into $J=1/2$, $J=3/2$ and $J=5/2$ sublocks, with the latter being the identity for all exchanges. Writing the unitaries in the basis states defined in \textbf{Appendix \ref{sec:AppStates}}, we obtain
\begin{align}
U^{1/2}_{12}(\Theta)=e^{i\Theta}
&\begin{bmatrix}
\frac{1}{4} & -\frac{\sqrt{3}}{4} & 0 & 0 & 0 \\
-\frac{\sqrt{3}}{4} & \frac{3}{4} & 0 & 0 & 0 \\
0 & 0 & \frac{1}{4} & \frac{-\sqrt{3}}{4} & 0 \\
0 & 0 & \frac{-\sqrt{3}}{4} & \frac{3}{4} & 0 \\
0 & 0 & 0 & 0 & 0 \\
\end{bmatrix}\nonumber\\+
&\begin{bmatrix}
\frac{3}{4} & \frac{\sqrt{3}}{4} & 0 & 0 & 0 \\
\frac{\sqrt{3}}{4} & \frac{1}{4} & 0 & 0 & 0 \\
0 & 0 & \frac{3}{4} & \frac{\sqrt{3}}{4} & 0 \\
0 & 0 & \frac{\sqrt{3}}{4} & \frac{1}{4} & 0 \\
0 & 0 & 0 & 0 & 1
\end{bmatrix}
\end{align}
\begin{equation}
U^{1/2}_{23}(\Theta)=e^{i\Theta}
\begin{bmatrix}
1&0&0&0&0\\
0&0&0&0&0\\
0&0&1&0&0\\
0&0&0&0&0\\
0&0&0&0&0\\
\end{bmatrix}+
\begin{bmatrix}
0&0&0&0&0\\
0&1&0&0&0\\
0&0&0&0&0\\
0&0&0&1&0\\
0&0&0&0&1\\
\end{bmatrix}
\end{equation}
\begin{align}
U^{1/2}_{34}(\Theta)=\frac{e^{i\Theta}}{12}
&\begin{bmatrix}
3 & 0 & 0 & 3 & -3 \sqrt{2} \\
0 & 3 & 3 & -2 \sqrt{3} & -\sqrt{6} \\
0 & 3 & 3 & -2 \sqrt{3} & -\sqrt{6} \\
3 & -2 \sqrt{3} & -2 \sqrt{3} & 7 & -\sqrt{2} \\
-3 \sqrt{2} & -\sqrt{6} & -\sqrt{6} & -\sqrt{2} & 8\end{bmatrix}\nonumber\\+\frac{1}{12}
&\begin{bmatrix}
9 & 0 & 0 & -3 & 3 \sqrt{2} \\
0 & 9 & -3 & 2 \sqrt{3} & \sqrt{6} \\
0 & -3 & 9 & 2 \sqrt{3} & \sqrt{6} \\
-3 & 2 \sqrt{3} & 2 \sqrt{3} & 5 & \sqrt{2} \\
3 \sqrt{2} & \sqrt{6} & \sqrt{6} & \sqrt{2} & 4\end{bmatrix}
\end{align}
\begin{equation}
U^{1/2}_{45}(\Theta)=e^{i\Theta}
\begin{bmatrix}
1&0&0&0&0\\
0&1&0&0&0\\
0&0&0&0&0\\
0&0&0&0&0\\
0&0&0&0&0\\
\end{bmatrix}+
\begin{bmatrix}
0&0&0&0&0\\
0&0&0&0&0\\
0&0&1&0&0\\
0&0&0&1&0\\
0&0&0&0&1\\
\end{bmatrix}
\end{equation}
for $J=1/2$ and
\begin{align}
U^{3/2}_{12}(\Theta)=e^{i\Theta}
\begin{bmatrix}
0 & 0 & 0 & 0 \\
0 & \frac{1}{4} & \frac{\sqrt{3}}{4} & 0 \\
0 & \frac{\sqrt{3}}{4} & \frac{3}{4} & 0 \\
0 & 0 & 0 & 0
\end{bmatrix}+
\begin{bmatrix}
1 & 0 & 0 & 0 \\
0 & \frac{3}{4} & -\frac{\sqrt{3}}{4} & 0 \\
0 & -\frac{\sqrt{3}}{4} & \frac{1}{4} & 0 \\
0 & 0 & 0 & 1\end{bmatrix}
\end{align}
\begin{align}
U^{3/2}_{23}(\Theta)=e^{i\Theta}
\begin{bmatrix}
0&0&0&0\\
0&1&0&0\\
0&0&0&0\\
0&0&0&0\\
\end{bmatrix}+
\begin{bmatrix}
1&0&0&0\\
0&0&0&0\\
0&0&1&0\\
0&0&0&1\\
\end{bmatrix}
\end{align}
\begin{align}
U^{3/2}_{34}(\Theta)=\frac{e^{i\Theta}}{12}
&\begin{bmatrix}
3 & 3 & -\sqrt{3} & \sqrt{15} \\
3 & 3 & -\sqrt{3} & \sqrt{15} \\
-\sqrt{3} & -\sqrt{3} & 1 & -\sqrt{5} \\
\sqrt{15} & \sqrt{15} & -\sqrt{5} & 5\end{bmatrix}\nonumber\\+\frac{1}{12}&\begin{bmatrix}
9 & -3 & \sqrt{3} & -\sqrt{15} \\
-3 & 9 & \sqrt{3} & -\sqrt{15} \\
\sqrt{3} & \sqrt{3} & 11 & \sqrt{5} \\
-\sqrt{15} & -\sqrt{15} & \sqrt{5} & 7\end{bmatrix}
\end{align}
\begin{align}
U^{3/2}_{45}(\Theta)=e^{i\Theta}
\begin{bmatrix}
1&0&0&0\\
0&0&0&0\\
0&0&0&0\\
0&0&0&0\\
\end{bmatrix}+
\begin{bmatrix}
0&0&0&0\\
0&1&0&0\\
0&0&1&0\\
0&0&0&1\\
\end{bmatrix}
\end{align}
for $J=3/2$.
Exchanges $J^{12}$ and $J^{23}$ generate the usual Exchange Only qubit operations, while exchange $J^{45}$ performs a $Z$-operation on the QA qubit. Exchange $J^{34}$ generates the needed interaction between different $j_\mathrm{Q}$ and $j_\mathrm{A}$ representations, allowing for the actual removal of leakage.
With the trivial evolution of the $J=5/2$ space, the complete unitaries are given by $U_{ij}(\Theta)=U_{ij}^{1/2}(\Theta)\oplus U_{ij}^{3/2}(\Theta)\oplus I^{5/2}$. Clearly, exchanges acting on disjoint spin pairs commute ($[U_{ij}(\Theta),U_{mn}(\Theta')]=0$ for $j\neq m$ and $i\neq n$ with the convention $i<j$ and $n<m$). With the chosen brickwork gate layout, the total global unitary is then given by
\begin{eqnarray}
U(\boldsymbol{\Theta})=&&U_{23}(\Theta_{20})U_{45}(\Theta_{19})U_{12}(\Theta_{18})U_{34}(\Theta_{17})\nonumber\\ &&\dots U_{12}(\Theta_2)U_{34}(\Theta_1).\label{seqq}\end{eqnarray}
Note, however, that $\Theta_{19}=0$ always, because this exchange operation is a place holder, present only to allow a systematic numbering of the brickwork. 

Finally, since we initialize the ancilla in a singlet state, the isometry under study is 
\begin{equation}
T(\boldsymbol{\Theta})=U(\boldsymbol{\Theta})|S_\mathrm{A}\rangle\langle S_\mathrm{A}|,\label{isom}
\end{equation} 
The projector $|S_\mathrm{A}\rangle\langle S_\mathrm{A}|$ does not commute with $U_{34}(\Theta_1)$ if it acts non-trivially, so only the unitaries in the first brickwork layer are reduced to isometries.

\section{Target function and numerical search procedure}
\label{sec:AppNumercialSearch}

In this Appendix we first derive the target function.  This function, $f^{tot}_{RiL}$ below, achieves its minimum value, $f^0_{RiL}=0$, only if the sequence of exchanges Eq. (\ref{seqq}) perform the reset-if-leaked functionality. 
To obtain the target function, we first write the general form of the desired reset-if-leaked gate isometry (cf. Eq. (\ref{isom})):  

\begin{align}
\hat{T}_\text{RiL} =\overbrace{\hat{U}_\mathrm{Q}\otimes\hat{T}_\text{QA}}^{J=1/2}\oplus\overbrace{\left(\alpha|6\rangle+\beta|7\rangle\right)\langle 5|}^{J=3/2}
\label{eq:T_RIL_Rep_Basis}
\end{align}
Here we use the states $|0\rangle-|8\rangle$ of Eqs. (\ref{A1}-\ref{A8}) of Appendix \ref{sec:AppStates}, and the tensor-product structure of the $J=1/2$ sector {\large\ding{182}} of Eq. (\ref{eq:Direct_Sum_structure_space}).

Note that the reset states $\{|6\rangle,|7\rangle\}$ span the Hilbert space of $\big(\mathcal{D}^\text{Q}_{1/2}\otimes\mathcal{D}^\text{A}_{1}\big)_{\frac{3}{2}}$, where discarding/measuring the ancillary spins may at maximum lead to gauge fixing (cf. Appendix \ref{sec:AppCohLeakage}).
It is the singlet-initialization requirement \ref{num:1_ancillaInitSinglet}, that causes Eq. \ref{eq:T_RIL_Rep_Basis} to be an isometry and not a full unitary operator, since it is only necessary to consider $|0\rangle$, $|1\rangle$ and $|5\rangle$ as initial states. Requirement \ref{num:4_donotInduceLeakage} means that two of the states in {\large\ding{185}}, $|4\rangle$ and $|8\rangle$, are not valid outcomes. \ref{num:2_donotEntangle} is enforced via the tensor product structure $\hat{U}_\mathrm{Q}\otimes\hat{T}_\text{QA}$. Finally the actual leakage removal requirement \ref{num:3_resetLeakedState} is carried out by transferring the excess angular momentum from the qubit space Q to the ancillary space QA, i.e. $\left(\alpha|6\rangle+\beta|7\rangle\right)\langle 5|$.

The optional requirements \ref{numopt:5_makeQubitGateNice} and \ref{numopt:6_makeSequenceFlaggable} are satisfied by a suitable restriction of the gates $\hat{U}_\mathrm{Q}$ and $\hat{T}_\text{QA}$, respectively. See the discussion starting at Eq. (\ref{eq:targetFunctionIdentity}).

To go further, we use our explicit tensor-product basis for the $J=1/2$ states, sector {\large\ding{182}}:
\begin{alignat}{3}
\big(\mathcal{D}_{1/2}^\text{Q1}\otimes\mathcal{D}_{0}^\text{Q23}\big)_{\frac{1}{2}}&\otimes\mathcal{D}^\text{A}_{0}&&\equiv |{0}_\mathrm{Q}\rangle\otimes|{0}_\mathrm{QA}\rangle, \\
\big(\mathcal{D}_{1/2}^\text{Q1}\otimes\mathcal{D}_{1}^\text{Q23}\big)_{\frac{1}{2}}&\otimes\mathcal{D}^\text{A}_{0}&&\equiv |{1}_\mathrm{Q}\rangle\otimes|{0}_\mathrm{QA}\rangle,\\
\big(\mathcal{D}_{1/2}^\text{Q1}\otimes\mathcal{D}_{0}^\text{Q23}&\otimes\mathcal{D}^\text{A}_{1}\big)_{\frac{1}{2}}&&\equiv |{0}_\mathrm{Q}\rangle\otimes|{1}_\mathrm{QA}\rangle,\\
\big[\big(\mathcal{D}_{1/2}^\text{Q1}\otimes\mathcal{D}_{1}^\text{Q23}\big)_{\frac{1}{2}}&\otimes\mathcal{D}^\text{A}_{1}\big]_{\frac{1}{2}}&&\equiv |{1}_\mathrm{Q}\rangle\otimes|{1}_\mathrm{QA}\rangle.
\end{alignat}
The flaggability condition \ref{numopt:6_makeSequenceFlaggable} is imposed via setting 
$\hat{T}_\text{QA}= |{0}_\mathrm{QA}\rangle \langle{0}_\mathrm{QA}|$ \bibnote{A measurement of the flag qubit in the $\{|{0}_\mathrm{QA}\rangle,|{1}_\mathrm{QA}\rangle\}$-basis corresponds to the outcome of a singlet-triplet ($\{S_\mathrm{A},T_\mathrm{A}\}$, total angular momentum) measurement in the ancilla dots, as the QA qubit is just a construction equivalent to the EO qubit using the gauge \cite{Kribs2005, Kribs2006} of the computational EO qubit as a third spin-1/2 component. The measurement therefore leaves the logical qubit invariant but may modify the computational gauge, which should be unproblematic -- but see Appendix \ref{sec:AppCohLeakage}}. If we do not care about flaggablity, we may allow the QA isometry all of its two degrees of freedom $(\varphi,\gamma)$, which end up as additional components for the optimization vector for the target function besides $\boldsymbol{\Theta}$:
\begin{align}
\hat{T}_\text{QA}(\varphi,\gamma)= \big(\cos(\gamma/2)|{0}_\mathrm{QA}\rangle+e^{i\varphi}\sin(\gamma/2)|{1}_\mathrm{QA}\rangle\big)\langle{0}_\mathrm{QA}|\label{angles}
\end{align}
The isometry on which the target function is built then reverses the ancillary isometry, so that $\hat{T}^0_\text{RiL}$ always outputs $|{0}_\mathrm{QA}\rangle$ in $\mathcal{H}_\mathrm{QA}$:
\begin{equation}
\hat{T}^0_\text{RiL}(\boldsymbol{\Theta},\varphi,\gamma) = \left[I_\mathrm{Q}\otimes\hat{T}_\text{QA}^\dagger(\varphi,\gamma)\oplus I_\frac{3}{2}\right]\hat{T}_\text{RiL}(\boldsymbol{\Theta}),
\end{equation}

The target function to be minimized can then be written as the sum of a suitable sum of powers (here chosen as two) of the components of the isometry which are supposed to vanish:
\begin{align}
f^0_\text{RiL}(\boldsymbol{\Theta},\varphi,\gamma) &= \sum_{i=2,3,4}\sum_{j=0,1}|\langle i|\hat{T}^0_\text{RiL}| j\rangle|^2\\
&+|\langle 5|\hat{T}^0_\text{RiL}| 5\rangle|^2+|\langle 8|\hat{T}^0_\text{RiL}| 5\rangle|^2.
\end{align}

If full flaggability is desired, the parameters $(\varphi,\gamma)$ in (\ref{angles}) are not included in the solution vector, thus they are not varied in the search procedure (e.g., set to zero). 

If a specific type of computational qubit gate is desired, we add a supplementary term to the target function $f_{U_Q}$, according to the gate $U_Q$ desired:
\begin{align}
f_\mathrm{Identity} &= 1-\left|\frac{1}{2}\mathrm{Tr}_\mathrm{Q}[\hat{U}_\mathrm{Q}]\right|\label{eq:targetFunctionIdentity}\\
f_\mathrm{Pauli} &= 1-\left\| \boldsymbol{c}\right\|_4 \label{eq:targetFunctionPauli}\\
f_\mathrm{Clifford} &= 2-\left\| \boldsymbol{c}^X\right\|_4-\left\| \boldsymbol{c}^Z\right\|_4\label{eq:targetFunctionClifford}
\end{align}
Here, $\left\| \cdot\right\|_4$ is the 4-norm, and $\boldsymbol{c}, \boldsymbol{c}^X, \boldsymbol{c}^Z \in \mathbb{C}^4$ have the components 
\begin{align}
c_n&=\frac{1}{2}\text{Tr}_\mathrm{Q}[\hat{U}_\mathrm{Q}P_n]\\
c_n^X&=\frac{1}{2}\text{Tr}_\mathrm{Q}[\hat{U}_\mathrm{Q} X \hat{U}_\mathrm{Q}^\dagger P_n],\\
c_n^Z&=\frac{1}{2}\text{Tr}_\mathrm{Q}[\hat{U}_\mathrm{Q} Z \hat{U}_\mathrm{Q}^\dagger P_n],
\end{align}
with $P_n\in\{I,X,Y,Z\}$ the set of Pauli gates. Then the total target function is 
\begin{equation}
f^\mathrm{tot}_\text{RiL}(\boldsymbol{\Theta},\varphi,\gamma)=f^\mathrm{tot}_\text{RiL}(\boldsymbol{\Theta},\varphi,\gamma)+f_{U_\mathrm{Q}}(\boldsymbol{\Theta},\varphi,\gamma). 
\end{equation}

The optimization algorithm of choice then aims to minimize $f^\mathrm{tot}_\text{RiL}(\boldsymbol{\Theta},\varphi,\gamma)$. A valid solution corresponds to $f^\mathrm{tot}_\text{RiL}(\boldsymbol{\Theta},\varphi,\gamma)=0$.
To implement the search procedure, we used a basin-hopping approach to avoid trapping of the search in local minima using the scipy implementation \cite{2020SciPy} based on the algorithm presented in \cite{Wales1997}. The underlying minimization routine was the scipy implementation of the Broyden-Fletcher-Goldfarb-Shanno algorithm \cite{Broyden1970, Fletcher1970, Goldfarb1970, Shanno1970}.
 The choice of parameters for the basin-hopping procedure were a temperature $T=10^{-5}$, a stepsize of $2\pi$, 100 iterations and an interval of 50. 

\section{Specific solution sequences}	
\label{sec:AppSolutions}

\begin{figure}%[ht!]
	%\centering
	\includegraphics[scale=0.5]{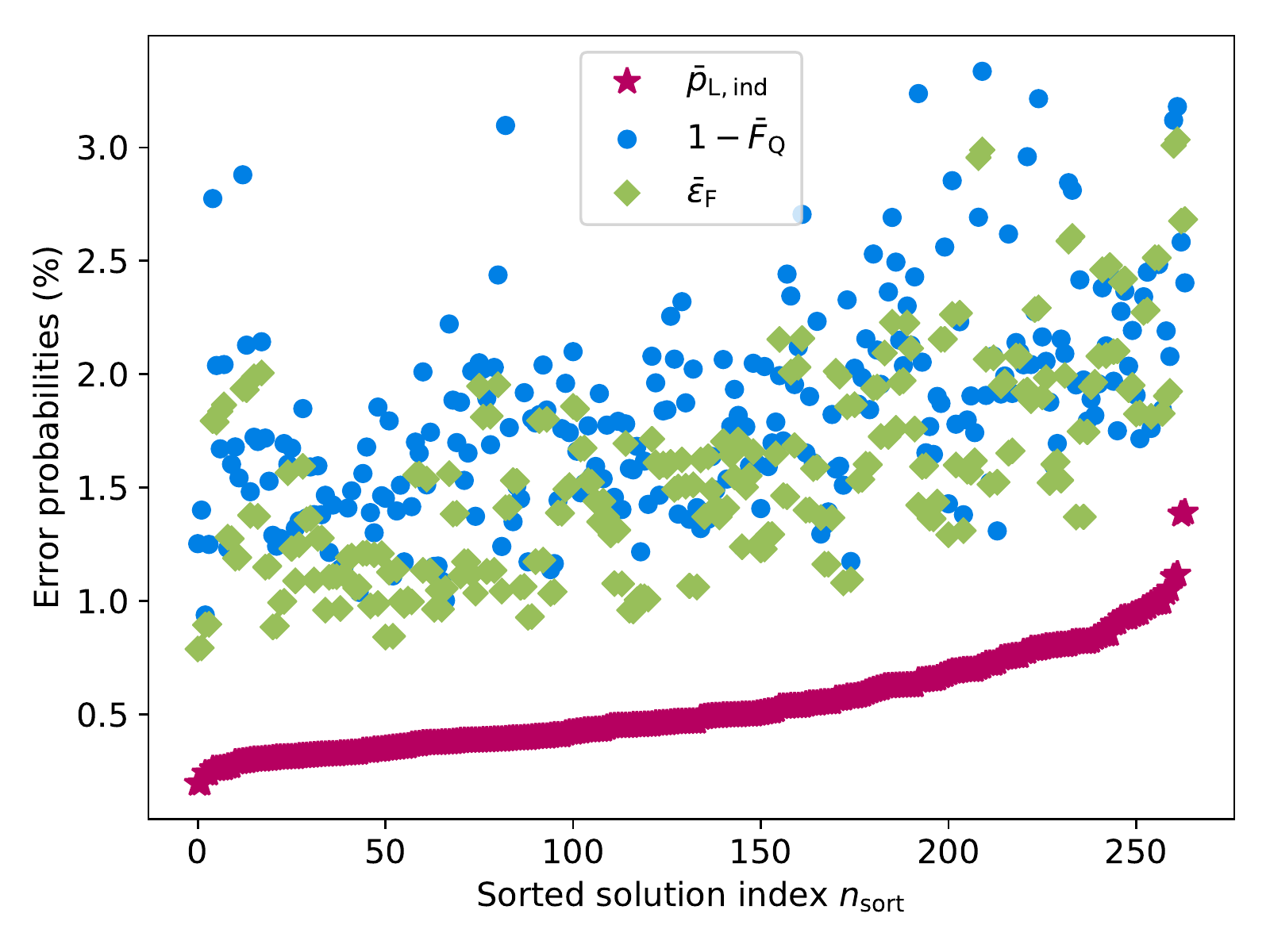}
	\includegraphics[scale=0.5]{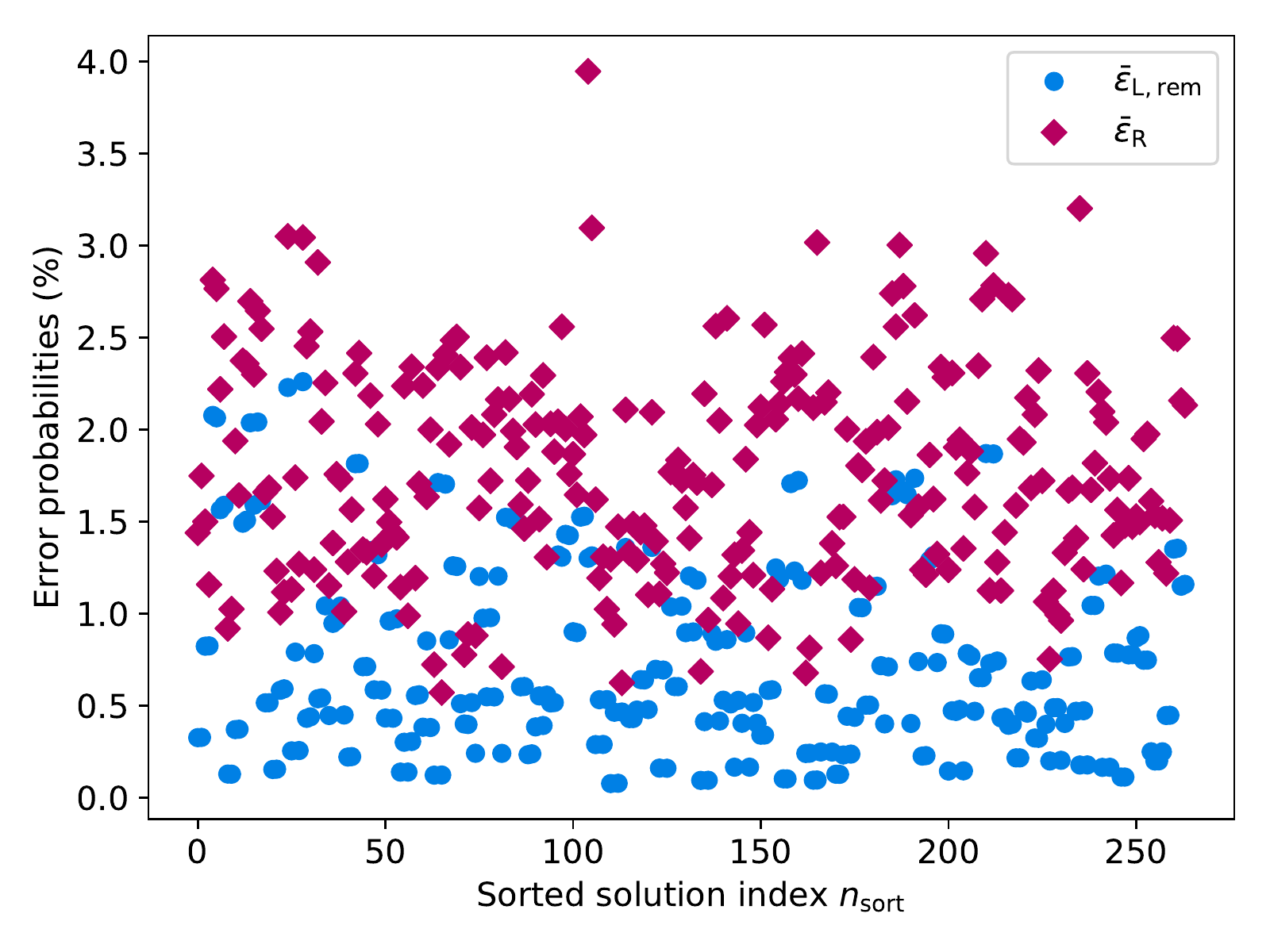}
	\caption{ Plot of the average error rates for a noise strength of $\sigma = 2\%$ for the 264 unique flaggable solutions found for a unleaked (top) and leaked (bottom) input qubit. The solutions are sorted by the average leakage induction $\bar{p}_\mathrm{L,ind}$.}
	\label{fig:AppRobustnessOfLeakageRemoval}
	\end{figure}

\begin{figure}%[ht!]
		%\centering
		\includegraphics[scale=0.5]{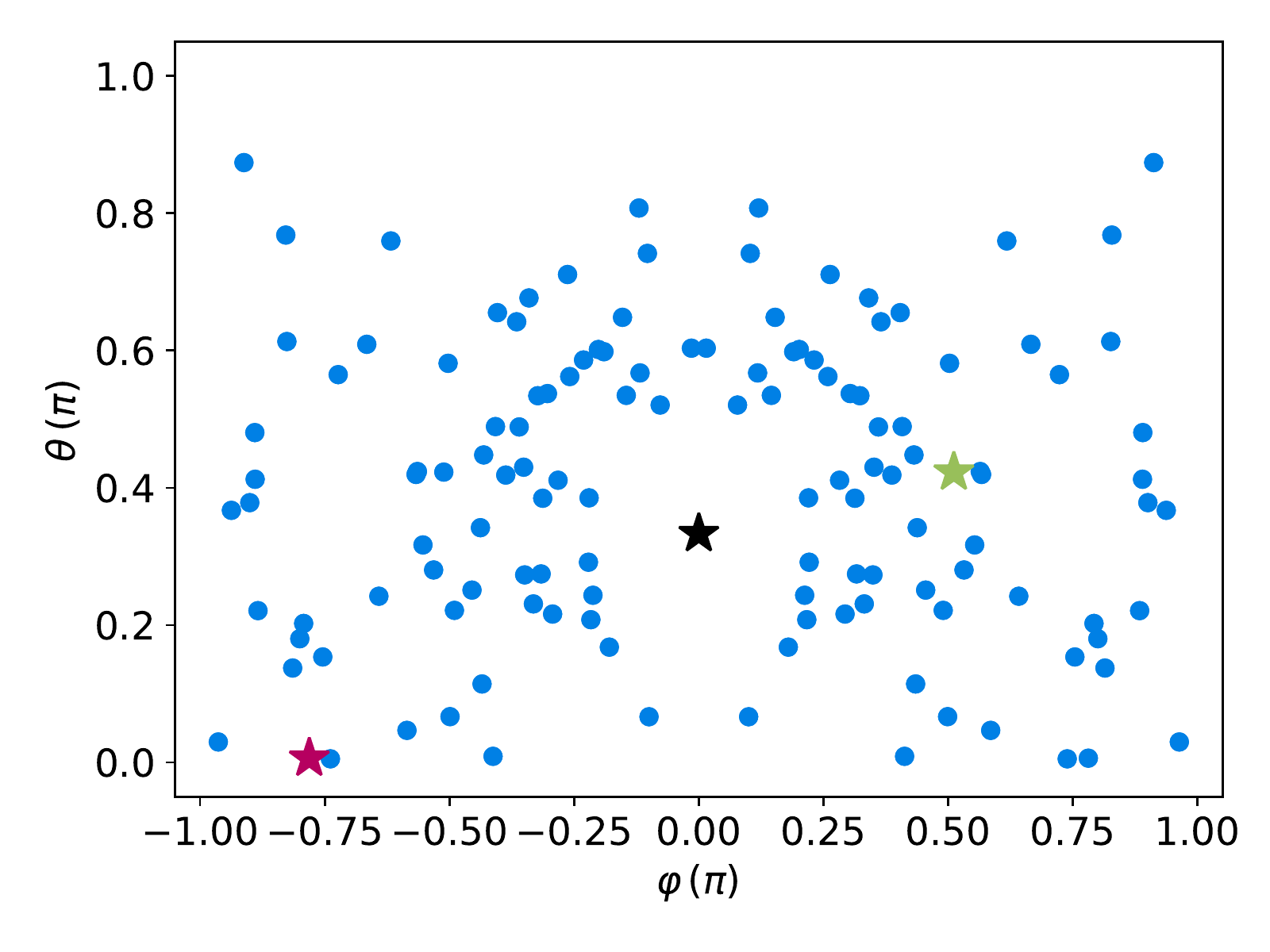}
		\caption{ Plot of the distribution of reset states, parameterized as $|\Psi_\mathrm{R}\rangle=\cos(\theta/2)|0_\mathrm{Q}\rangle+e^{i\varphi}\sin(\theta/2)|1_\mathrm{Q}\rangle$. Marked with stars are the the best (green) and worst (red) flaggable solutions, as well as the non-flaggable solution (black), see \textbf{Table \ref{tab:AppExchangeTable}}.}
		\label{fig:flagSolutionsResetStates}
	\end{figure}

	Table \ref{tab:AppExchangeTable} lists the solutions investigated for varying noise strengths in the main text. As mentioned there, the flaggable solution allowed for 264 distinct solutions.
	\begin{table}
		\centering
		\begin{tabular}{c@{\hskip 10pt}c@{\hskip 10pt}r@{\hskip 10pt}r}
			$\Theta_i$ & no flag & best flag & worst flag \\
			\hline
			$\textcolor{IYellow}{\Theta_1}$ & $q_1$ & 0.496474 & 1.540024 \\
			$\textcolor{QBlue}{\Theta_2}$ & $0$ & 0.511053 & 1.988000 \\
			$\textcolor{ARed}{\Theta_3}$ & $2-q_1$ & 0.407919 & 1.646738 \\
			$\textcolor{QBlue}{\Theta_4}$ & $1$ & 1.128462 & 0.463540 \\
			$\textcolor{IYellow}{\Theta_5}$ & $3/2$ & 0.644573 & 1.603884 \\
			$\textcolor{QBlue}{\Theta_6}$ & $3/2$ & 1.456051 & 0.829024 \\
			$\textcolor{ARed}{\Theta_7}$ & $0$ & 0.233065 & 1.183458 \\
			$\textcolor{QBlue}{\Theta_8}$ & $1$ & 1.473077 & 1.404117 \\
			$\textcolor{IYellow}{\Theta_9}$  & $3/2$ & 1.574455 & 0.613810 \\
			$\textcolor{QBlue}{\Theta_{10}}$ & $1/2$ & 1.481738 & 1.416749 \\
			$\textcolor{ARed}{\Theta_{11}}$ & $0$ & 0.296057 & 1.310604 \\
			$\textcolor{QBlue}{\Theta_{12}}$ & $1$ & 0.778243 & 1.379647 \\
			$\textcolor{IYellow}{\Theta_{13}}$ & $3/2$ & 0.458866 & 1.556976 \\
			$\textcolor{QBlue}{\Theta_{14}}$ & $3/2$ & 0.762262 & 1.310274 \\
			$\textcolor{ARed}{\Theta_{15}}$ & $1$ & 0.654983 & 0.517602 \\
			$\textcolor{QBlue}{\Theta_{16}}$ & $1$ & 0.907327 & 1.411259 \\
			$\textcolor{IYellow}{\Theta_{17}}$ & $1$ & 0.495382 & 1.144766 \\
			$\textcolor{QBlue}{\Theta_{18}}$ & $0$ & 0.403991 & 0.015345 \\
			$\textcolor{Grey}{\Theta_{19}}$ & \textcolor{Grey}{0} & \textcolor{Grey}{0} & \textcolor{Grey}{0} \\
			$\textcolor{QBlue}{\Theta_{20}}$ & $0$ & 1.700957 & 0.516077 \\
		\end{tabular}
		\caption{Exchange values for the studied sequences in units of $\pi$. $\Theta_{19}$ is included for completeness but may always be set to zero, as it acts either before ancilla disposal or singlet-triplet measurement.}
		\label{tab:AppExchangeTable}
	\end{table}
	Fig.~\ref{fig:AppRobustnessOfLeakageRemoval} (top) explores the properties of induced leakage probability (Eq.~(\ref{firsterror})), qubit error (Eq.~(\ref{seconderror})) and flaggability error (Eq.~(\ref{thirderror})) for a noise strength of $\sigma=2\%$, while Fig.~\ref{fig:AppRobustnessOfLeakageRemoval} (bottom) investigates the probabilities of failing to remove leakage and failing to end up in the desired reset state; these are given as
	\begin{align}
	\bar{\epsilon}_\mathrm{L,rem} &= \langle 5|\mathcal{E}(|5\rangle\langle5|)|5\rangle+\langle 8|\mathcal{E}(|5\rangle\langle5|)|8\rangle\\
	\bar{\epsilon}_\mathrm{R} &= 1-\langle \Psi_\mathrm{R}|\mathcal{E}(|5\rangle\langle 5|)|\Psi_\mathrm{R}\rangle
	\end{align}
	with $\bar{\epsilon}_\mathrm{L,rem} \leq \bar{\epsilon}_\mathrm{R}$: If the qubit were initially leaked, then we may either fail to remove leakage, or return it to a different reset state from the chosen one. The latter still accomplishes the goal of leakage removal, making it a low priority error. Since we assume the initial leakage population to be small, the above errors are of second order importance.
	Finally, Fig.~\ref{fig:flagSolutionsResetStates} plots the reset states $|\Psi_\mathrm{R}\rangle=\cos(\theta/2)|0_\mathrm{Q}\rangle+e^{i\varphi}\sin(\theta/2)|1_\mathrm{Q}\rangle$ for the different sequences.

\section{Noise Study Details}
\label{sec:AppNoiseStudyDetails}

The starting point for the noise study is Eq. (\ref{eq:noisyChannelGeneral}) from the main text:
\begin{equation}
\mathcal{E}(\rho) = \int \mathrm{d}\boldsymbol{x} \, p(\boldsymbol{x})\,\hat{T}[\boldsymbol{\Theta}(\boldsymbol{x})]\rho \hat{T}^\dagger[\boldsymbol{\Theta}(\boldsymbol{x})].
\end{equation}
We first decompose the general isometry $\hat{T}(\boldsymbol{\Theta})$ into the total spin angular momentum subblocks:
\begin{equation}
\hat{T}(\boldsymbol{\Theta})=\hat{T}^{1/2}(\boldsymbol{\Theta})\oplus\hat{T}^{3/2}(\boldsymbol{\Theta})\oplus 0^{5/2}.
\end{equation}
These subblocks are expressed in the basis states defined in \textbf{Appendix \ref{sec:AppStates}}:
\begin{align}
T^{1/2}=\begin{matrix}&\langle 0| \quad\langle 1|\quad\\
\begin{matrix}
|0\rangle\\
|1\rangle\\
|2\rangle\\
|3\rangle\\
|4\rangle\\
\end{matrix}
&\begin{bmatrix}
T_{00} & T_{01}\\
T_{10} & T_{11}\\
T_{20} & T_{21}\\
T_{30} & T_{31}\\
T_{40} & T_{41}\\
\end{bmatrix}
\end{matrix}
\end{align}

\begin{align}
T^{3/2}=\begin{matrix}&\langle 5|\\
\begin{matrix}
|5\rangle\\
|6\rangle\\
|7\rangle\\
|8\rangle\\
\end{matrix}
&\begin{bmatrix}
T_{55} \\
T_{65} \\
T_{75} \\
T_{85} \\
\end{bmatrix}
\end{matrix}
\end{align}

In terms of these submatrices, the $14\times14$ $\chi$ matrix of a single channel realization is straightforwardly given by the outer product of the vector
\begin{equation}
\boldsymbol{v} =
\begin{bmatrix}
\mathrm{vec}(T^{1/2})\\T^{3/2}
\end{bmatrix}
\end{equation}
(with $\mathrm{vec}(\cdot)$ indicating matrix vectorization) as $\chi = \boldsymbol{v} \boldsymbol{v}^\dagger$
in the operator basis
\begin{align} 
E_j\in\{&|0\rangle\langle 0|,|1\rangle\langle 0|,|2\rangle\langle 0|,|3\rangle\langle 0|,|4\rangle\langle 0|,\nonumber\\&|0\rangle\langle 1|,|1\rangle\langle 1|,|2\rangle\langle 1|,|3\rangle\langle 1|,|4\rangle\langle 1|,\nonumber\\&|5\rangle\langle 5|,|6\rangle\langle 5|,|7\rangle\langle 5|,|8\rangle\langle 5|\}.
\end{align}
Because this basis is independent of $\boldsymbol{\Theta}$, the total channel average can be written as 
\begin{align}
\mathcal{E}(\boldsymbol{\Theta}_0,\rho) &= \int \mathrm{d}\boldsymbol{x}\,p(\boldsymbol{x})\mathcal{E}_0[\boldsymbol{\Theta}(\boldsymbol{x}),\rho] \nonumber\\&= \sum_{ij} E_i\rho E_j \int \mathrm{d}\boldsymbol{x}\,p(\boldsymbol{x})\chi_{ij}[\boldsymbol{\Theta}(\boldsymbol{x})]\nonumber\\&\equiv \sum_{ij} E_i\rho E_j \bar{\chi}_{ij}[\boldsymbol{\Theta}_0,p(\boldsymbol{x})].
\end{align}
Recalling that $\Theta_i(\boldsymbol{x}) = \Theta_{0,i}\times(1+x_i)$, the matrix $\bar{\chi}[\boldsymbol{\Theta}_0,p(\boldsymbol{x})]$ then contains all the average channel properties for the channel which would ideally be realized via the exchange sequence $\boldsymbol{\Theta}_0$.
$\bar{\chi}$ was numerically sampled using the distribution $p(\boldsymbol{x})$ in Eq. (\ref{eq:fullNonMarkovModel})  for the non-Markovian noise model from the main text (switching to the fully Markovian case did not influence the mean significantly), using 100,000 samples per average, yielding a relative standard error of the mean of $<3.5\times10^{-3}$ for all estimators.

\section{Coherence of Leakage and Gauge behavior}
\label{sec:AppCohLeakage}
This appendix considers the conversion and generation of coherent errors by the RiL procedure investigated in the paper.

Coherent errors are viewed as disadvantageous compared to stochastic ones, due to potentially stronger error build up under repeated application of gates giving rise to such errors \cite{Sanders2015}. We therefore discuss the coherence properties of both leakage production as well as reset of leakage components in coherent superposition with the unleaked states.

The answer is fairly simple for the flaggable solution: An intially unleaked state will be accompanied by a singlet in the ancillary spins, while both induced leakage as well as reset states will be entangled with an ancillary triplet, which means that discarding the ancillary spins (i.e, releasing them into the environment) will decohere any superposition between the initally unleaked and the induced leakage and reset states. The unflaggable case, however, requires somewhat more discussion, as all the ancillary output states will have triplet character, so our argument is not as straightforward.

First, we deal with the case of resetting a potentially coherent leakage state. 
The origin of such coherence may be rooted in other gates or unwanted physical actions (e.g. by inter-dot magnetic field gradients, but the detailed mechanisms are outside the scope of this discussion) that may generate coherent leakage. We argue below that the RiL procedure does not. Further, we discuss here if a coherent leakage state lives on as a coherent error after reset.

For the question of coherence of reset states, we may fix the gauge (i.e., the $m$ quantum number) for now. We will consider a general computational state $|\psi_\mathrm{U}\rangle$ in coherent superposition (see Eqs. (\ref{A10}-\ref{A11})) with a leakage state $|\mathrm{L}\rangle$ compatible with its gauge ($J_+$ applied to Eq. (\ref{A15})):
\begin{equation}
\left(\alpha|\psi_\mathrm{U}\rangle+\beta|\mathrm{L}\rangle\right)\otimes|S\rangle_\mathrm{A}.
\end{equation}
(If the gauges aren't compatible, errors will automatically be incoherent.) The perfect RiL procedure will transfer the support of $|\psi_\mathrm{U}\rangle$ from Eqs. (\ref{A10}-\ref{A11}) to Eqs. (\ref{A12}-\ref{A13}) and the $|\mathrm{L}\rangle$ state into the reset state $|\psi_\mathrm{R}\rangle$ (spanned by Eqs. (\ref{A16}-\ref{A17}) with the right number of $J_+$s applied), and will, in general, entangle the gauge with the ancillary state.
Setting the incoming gauge to $m=\,\downarrow$ without loss of generality, the outgoing state will then be:
\begin{align}
&\sqrt{\frac{1}{3}}\left[ \left( \alpha|\psi_\mathrm{U}\rangle+\sqrt{2}\beta|\psi_\mathrm{R}\rangle\right)|\downarrow T_0\rangle \right.\nonumber \\&\quad-\left.\left(\sqrt{2}\alpha|\psi_\mathrm{U}\rangle-\beta|\psi_\mathrm{R}\rangle\right)|\uparrow T_-\rangle\right]
\end{align}
where we explicity write out the resulting gauge state as well as the triplet state of the ancillary spins. As we are going to discard the ancillary states in the next step, this already indicates the upcoming gauge mixing due to the RiL procedure. Explicitly written out, we have the outgoing density matrix
\begin{widetext}
\begin{align}
&\frac{1}{3}\left[\left(|\alpha|^2|\psi_\mathrm{U}\rangle\langle\psi_\mathrm{U}|+2|\beta|^2|\psi_\mathrm{R}\rangle\langle\psi_\mathrm{R}|+\sqrt{2}\left(\alpha\beta^*|\psi_\mathrm{U}\rangle\langle\psi_\mathrm{R}|+\alpha^*\beta|\psi_\mathrm{R}\rangle\langle\psi_\mathrm{U}|\right)\right)|\downarrow\rangle\langle\downarrow|\otimes|T_0\rangle\langle T_0|\right.\nonumber\\&\left.\,+\left(2|\alpha|^2|\psi_\mathrm{U}\rangle\langle\psi_\mathrm{U}|+|\beta|^2|\psi_\mathrm{R}\rangle\langle\psi_\mathrm{R}|-\sqrt{2}\left(\alpha\beta^*|\psi_\mathrm{U}\rangle\langle\psi_\mathrm{R}|+\alpha^*\beta|\psi_\mathrm{R}\rangle\langle\psi_\mathrm{U}|\right)\right)|\uparrow\rangle\langle\uparrow|\otimes|T_-\rangle\langle T_-|\right]
\end{align}
\end{widetext}
where we already dropped the off-diagonal terms. Discarding the ancillary states will now lead both to mixing of the gauge as well as canceling out the coherent terms between unleaked and reset state on average.
Keeping track of the gauge state separately, the output density matrix for the qubit state will then be
\begin{align}
&|\alpha|^2|\psi_\mathrm{U}\rangle\langle\psi_\mathrm{U}|+|\beta|^2|\psi_\mathrm{R}\rangle\langle\psi_\mathrm{R}|
\end{align}
lacking any coherence between unleaked state and reset state.

Second, we deal with the question of coherent creation of leakage due to the procedure, e.g. caused, as described in the main text, by charge noise with long correlation times.

We explicitly write out the representations of the $J=1/2$ sector. Due to initialization in  $\mathcal{D}^\text{QA}_{1/2}=\mathcal{D}^\text{Q}_{1/2}\otimes\mathcal{D}^\text{A}_{0}$, the global magnetic quantum number is fixed to the initial value of the qubit gauge, i.e. $M=m\in\{\downarrow,\uparrow\}$.
The internal structure of $\mathcal{D}^\text{Q}_{1/2}$ is not of concern here, but note that it may end up different in the two outputs $\mathcal{D}^\text{Q}_{1/2}\otimes\mathcal{D}^\text{A}_{0}$ and $\left(\mathcal{D}^\text{Q}_{1/2}\otimes\mathcal{D}^\text{A}_{1}\right)_{1/2}$. In this case, discarding the ancillary spins leads to an incoherent error in a straightforward manner. 
In the following, we use $\{|-\tfrac{1}{2}_\mathrm{U}\rangle,|+\tfrac{1}{2}_\mathrm{U}\rangle\}$ to refer to the spin-1/2 (unleaked) states $\{|j=1/2,m=+ 1/2\rangle,|j=1/2,m=-1/2\rangle\}$ and $\{|m_\mathrm{L}\rangle\}$ to refer to the $m$ projections of the $j=3/2$ (leaked) states. We write down the relevant states for the initial gauge $m=- 1/2$:
\begin{alignat}{3}
\left|{0}_\mathrm{QA},-\tfrac{1}{2}\right\rangle &= \left|-\tfrac{1}{2}_\mathrm{U}\right\rangle_\mathrm{Q}|S\rangle_\mathrm{A}\,\,\,{\mbox{(cf. Eqs. (\ref{A10}- \ref{A11}));}} \\
\left|{1}_\mathrm{QA},-\tfrac{1}{2}\right\rangle &= +\sqrt{\frac{1}{3}}\left|-\tfrac{1}{2}_\mathrm{U}\right\rangle_\mathrm{Q}|T_0\rangle_\mathrm{A}-\sqrt{\frac{2}{3}}\left|+\tfrac{1}{2}_\mathrm{U}\right\rangle_\mathrm{Q}|T_-\rangle_\mathrm{A} \\
&{\mbox{(cf. Eqs. (\ref{A12}- \ref{A13}));}} \nonumber\\
\left|{2}_\mathrm{Q},-\tfrac{1}{2}\right\rangle &= +\sqrt{\frac{1}{2}}\left|-\tfrac{3}{2}_\mathrm{L}\right\rangle_\mathrm{Q}|T_+\rangle_\mathrm{A}\,\,\,\,\,\,\,\,\,{\mbox{(cf. Eq. (\ref{A14}))}}\nonumber\\ &\;\;\;\;-\sqrt{\frac{1}{3}}\left|-\tfrac{1}{2}_\mathrm{L}\right\rangle_\mathrm{Q}|T_0\rangle_\mathrm{A}+\sqrt{\frac{1}{6}}\left|+\tfrac{1}{2}_\mathrm{L}\right\rangle_\mathrm{Q}|T_-\rangle_\mathrm{A}\nonumber\\
\end{alignat}
and $m=+1/2$ (with the same correspondence to Eqs. (\ref{A10}- \ref{A14})):
\begin{alignat}{3}
\left|{0}_\mathrm{QA},+\tfrac{1}{2}\right\rangle &= \left|+\tfrac{1}{2}_\mathrm{U}\right\rangle_\mathrm{Q}|S\rangle_\mathrm{A} \\
\left|{1}_\mathrm{QA},+\tfrac{1}{2}\right\rangle &= -\sqrt{\frac{1}{3}}\left|+\tfrac{1}{2}_\mathrm{U}\right\rangle_\mathrm{Q}|T_0\rangle_\mathrm{A}+\sqrt{\frac{2}{3}}\left|-\tfrac{1}{2}_\mathrm{U}\right\rangle_\mathrm{Q}|T_+\rangle_\mathrm{A} \\
\left|{2}_\mathrm{Q},+{\tfrac{1}{2}}\right\rangle &= +\sqrt{\frac{1}{2}}\left|+{\tfrac{3}{2}}_\mathrm{L}\right\rangle_\mathrm{Q}|T_-\rangle_\mathrm{A}\\ &\;\;\;\;-\sqrt{\frac{1}{3}}\left|+{\tfrac{1}{2}}_\mathrm{L}\right\rangle_\mathrm{Q}|T_0\rangle_\mathrm{A}+\sqrt{\frac{1}{6}}\left|-{\tfrac{1}{2}}_\mathrm{L}\right\rangle_\mathrm{Q}|T_+\rangle_\mathrm{A}\nonumber
\end{alignat}
Since we will discard the ancilla, we consider the effect of decohering in the ancilla basis 
$\{|S\rangle,|T_-\rangle,|T_0\rangle,|T_+\rangle\}$. The $|S\rangle$ component straightforwardly singles out the $|{0}_\mathrm{QA}\rangle$ block. There are, however, remnants of coherence between qubit and leakage space left after decohering the triplet components. Specifically, if our sequence were to end up in the state $\alpha|{1}_\mathrm{QA}\rangle+\beta|{2}_\mathrm{Q}\rangle$, and omitting all off-diagonal terms in the triplet basis, the output density matrix would be:
\begin{align}
\frac{1}{2}|&\beta|^2\left|-\tfrac{3}{2}_\mathrm{L}\right\rangle_\mathrm{Q}\left\langle-\tfrac{3}{2}_\mathrm{L}\right|_\mathrm{Q}\otimes|T_+\rangle_\mathrm{A}\langle T_+|_\mathrm{A}\\
+\frac{1}{3}&\bigg(|\alpha|^2\left|-\tfrac{1}{2}_\mathrm{U}\right\rangle_\mathrm{Q}\left\langle-\tfrac{1}{2}_\mathrm{U}\right|_\mathrm{Q}+\alpha\beta^*\left|-\tfrac{1}{2}_\mathrm{U}\right\rangle_\mathrm{Q}\left\langle-\tfrac{1}{2}_\mathrm{L}\right|_\mathrm{Q}\nonumber\\+\alpha^*\beta&\left|-\tfrac{1}{2}_\mathrm{L}\right\rangle_\mathrm{Q}\left\langle-\tfrac{1}{2}_\mathrm{U}\right|_\mathrm{Q}+|\beta|^2\left|-\tfrac{1}{2}\right\rangle_\mathrm{Q}\left\langle-\tfrac{1}{2}_\mathrm{L}\right|_\mathrm{Q}\bigg)\otimes|T_0\rangle_\mathrm{A}\langle T_0|_\mathrm{A}\nonumber\\
+\frac{1}{6}&\bigg(4|\alpha|^2\left|+\tfrac{1}{2}_\mathrm{U}\right\rangle_\mathrm{Q}\left\langle+\tfrac{1}{2}_\mathrm{U}\right|_\mathrm{Q}+2\alpha\beta^*\left|+\tfrac{1}{2}_\mathrm{U}\right\rangle_\mathrm{Q}\left\langle+\tfrac{1}{2}_\mathrm{L}\right|_\mathrm{Q}\nonumber\\+2\alpha^*\beta&\left|+\tfrac{1}{2}_\mathrm{L}\right\rangle_\mathrm{Q}\left\langle+\tfrac{1}{2}_\mathrm{U}\right|_\mathrm{Q}+|\beta|^2\left|+\tfrac{1}{2}_\mathrm{L}\right\rangle_\mathrm{Q}\left\langle+\tfrac{1}{2}_\mathrm{L}\right|_\mathrm{Q}\bigg)\otimes|T_-\rangle_\mathrm{A}\langle T_-|_\mathrm{A}\nonumber
\end{align}
for initial $m=-1/2$ and 
\begin{align}
\frac{1}{2}|&\beta|^2\left|+\tfrac{3}{2}_\mathrm{L}\right\rangle_\mathrm{Q}\left\langle+\tfrac{3}{2}_\mathrm{L}\right|_\mathrm{Q}\otimes|T_-\rangle_\mathrm{A}\langle T_-|_\mathrm{A}\\
+\frac{1}{3}&\bigg(|\alpha|^2\left|+\tfrac{1}{2}_\mathrm{U}\right\rangle_\mathrm{Q}\left\langle+\tfrac{1}{2}_\mathrm{U}\right|_\mathrm{Q}-\alpha\beta^*\left|+\tfrac{1}{2}_\mathrm{U}\right\rangle_\mathrm{Q}\left\langle+\tfrac{1}{2}_\mathrm{L}\right|_\mathrm{Q}\nonumber\\-\alpha^*\beta&\left|+\tfrac{1}{2}_\mathrm{L}\right\rangle_\mathrm{Q}\left\langle+\tfrac{1}{2}_\mathrm{U}\right|_\mathrm{Q}+|\beta|^2\left|+\tfrac{1}{2}\right\rangle_\mathrm{Q}\left\langle+\tfrac{1}{2}\right|_\mathrm{Q}\bigg)\otimes|T_0\rangle_\mathrm{A}\langle T_0|_\mathrm{A}\nonumber\\
+\frac{1}{6}&\bigg(4|\alpha|^2\left|-\tfrac{1}{2}_\mathrm{U}\right\rangle_\mathrm{Q}\left\langle-\tfrac{1}{2}_\mathrm{U}\right|_\mathrm{Q}-2\alpha\beta^*\left|-\tfrac{1}{2}_\mathrm{U}\right\rangle_\mathrm{Q}\left\langle-\tfrac{1}{2}_\mathrm{L}\right|_\mathrm{Q}\nonumber\\-2\alpha^*\beta&\left|-\tfrac{1}{2}_\mathrm{L}\right\rangle_\mathrm{Q}\left\langle-\tfrac{1}{2}_\mathrm{U}\right|_\mathrm{Q}+|\beta|^2\left|-\tfrac{1}{2}_\mathrm{L}\right\rangle_\mathrm{Q}\left\langle-\tfrac{1}{2}\right|_\mathrm{Q}\bigg)\otimes|T_+\rangle_\mathrm{A}\langle T_+|_\mathrm{A}\nonumber
\end{align}
for initial $m=+1/2$. The only information about our five-dot system that we abandon is the state of the ancilla spins, which, if there is no entanglement between Q and QA, does not lead to any effects beyond gauge fixing.

The isometry which restores us to the representation level may be written as
\begin{equation}
T_{\mathrm{ms}} = \sum_{i=0}^8|i\rangle\sum_{m_i} \langle i,m_i|,
\label{eq:AppGauge2Nongauge}
\end{equation}
grouping together all the spin projections belonging to the individual representations which behave the same under Heisenberg exchange listed in \textbf{Appendix \ref{sec:AppStates}}.

We may now carry out the trace and return to the representation level, simply by dropping the details of $m$ and only discriminating between $j=1/2$ and $j=3/2$. Substituting the unleaked qubit state $|\psi_\mathrm{U}\rangle$ for $|{1}_\mathrm{QA}\rangle$, we obtain
\begin{equation}
|\alpha|^2\left|\psi_\mathrm{U}\right\rangle\left\langle\psi_\mathrm{U}\right|+|\beta|^2\left|{2}_\mathrm{Q}\right\rangle\left\langle{2}_\mathrm{Q}\right|+\frac{2}{3}\left(\alpha\beta^*\left|\psi_\mathrm{U}\right\rangle\left\langle{2}_\mathrm{Q}\right|+{\mbox{h.c.}}\right)
\end{equation}
for initial $m=-1/2$ and
\begin{equation}
|\alpha|^2\left|\psi_\mathrm{U}\right\rangle\left\langle\psi_\mathrm{U}\right|+|\beta|^2\left|{2}_\mathrm{Q}\right\rangle\left\langle{2}_\mathrm{Q}\right|-\frac{2}{3}\left(\alpha\beta^*\left|\psi_\mathrm{U}\right\rangle\left\langle{2}_\mathrm{Q}\right|+{\mbox{h.c.}}\right)
\end{equation}
for initial $m=+1/2$.
While these two density matrices do not represent pure states, they do retain a non-negligible amount of coherence (minimal purity of $13/18$, in the unrealistic case that the RiL produces 50\% leakage). However, the off-diagonal components are of opposite sign, which means that after application of isometry (\ref{eq:AppGauge2Nongauge}) the coherence of the output state will depend on the state of the inital input gauge:
\begin{align}
&|\alpha|^2\left|\psi_\mathrm{U}\right\rangle\left\langle\psi_\mathrm{U}\right|+|\beta|^2\left|{2}_\mathrm{Q}\right\rangle\left\langle{2}_\mathrm{Q}\right|\\+\frac{2}{3}&\mathrm{Tr}_\mathrm{G}[\rho_\mathrm{G}Z_\mathrm{G}]\left(\alpha\beta^*\left|\psi_\mathrm{U}\right\rangle\left\langle{2}_\mathrm{Q}\right|+{\mbox{h.c.}}\right)\nonumber
\end{align}
Here, $\rho_\mathrm{G}$ is the density matrix of the gauge degree of freedom, and $Z_\mathrm{G}$ the corresponding Pauli $Z$ operator.
Under perfect operation, the unflaggable RiL procedure performs incoherent pumping of the gauge state with the doubly stochastic matrix
\begin{equation}
P_\mathrm{G}= \frac{1}{3}\begin{bmatrix}
1 & 2 \\
2 & 1 \\
\end{bmatrix}.
\end{equation}
Counteracting this pumping behavior is the natural spin-relaxation due to spin-orbit effects, which we will describe by a phenomenological stochastic matrix
\begin{equation}
R = \begin{bmatrix}
1 & \eta \\
0 & 1-\eta \\
\end{bmatrix}.
\end{equation}
The stationary state of $RP_\mathrm{G}$ is:
\begin{equation}
\begin{pmatrix}
p_{\downarrow,\mathrm{s}}\\
p_{\uparrow,\mathrm{s}}
\end{pmatrix}=
\frac{1}{2}\begin{pmatrix}
1+\frac{3\eta}{4-\eta}\\
1-\frac{3\eta}{4-\eta}
\end{pmatrix}
\end{equation}
The other eigenstate is $(-1,1)^\mathrm{T}$	with eigenvalue $-(1-\eta)/3$, indicative of its decay behavior.
The density matrix of the stationary gauge state is therefore
\begin{equation}
\rho_\mathrm{G} = \frac{I_\mathrm{G}}{2}+\frac{3\eta}{4-\eta}\frac{Z_\mathrm{G}}{2}\label{postpump}
\end{equation}
with the gauge identity operator $I_\mathrm{G}$.  Equation (\ref{postpump}) is the expected gauge state just before a RiL procedure a few cycles after "starting up the quantum computation".

The spin orbit effects leading to gauge relaxation $\eta$ are one of the processes leading to leakage (others being the ones discussed above). Since we wish to employ the RiL procedure for small leakage populations, the decay of the gauge should be similarly small, i.e. $\eta\ll1$, so that the coherent leakage terms are approximately weighted by
\begin{equation}
\mathrm{Tr}_\mathrm{G}[\rho_\mathrm{G}Z_\mathrm{G}]\approx 3\eta/4
\end{equation}
and should therefore be negligible.

In conclusion, production of leakage due to the RiL procedure becomes fully incoherent if the computational gauge is in a fully mixed state. A non-flaggable RiL procedure does naturally lead to such a gauge state even in the absence of e.g. gauge thermalization, so any leakage induced by the RiL procedure is incoherent.

\section{Error calculations for flag error discussion}
\label{sec:AppFlaggingConditionalProbabilites}

We first wish to compute the probability that the measurement result yields a false negative result, i.e. measuring a negative (indicating no leakage) flag result $0_\mathrm{M}$ in every case \textit{except} that the input was not leaked and the output is still unleaked:
\begin{align}
&P(\overline{U_\mathrm{out}U_\mathrm{in}}|0_\mathrm{M}) \nonumber\\&= P(L_\mathrm{out}U_\mathrm{in}|0_\mathrm{M})+P(U_\mathrm{out}L_\mathrm{in}|0_\mathrm{M})+P(L_\mathrm{out}L_\mathrm{in}|0_\mathrm{M})
\nonumber\\
&=\frac{1}{P(0_\mathrm{M})}\big[P(0_\mathrm{M}L_\mathrm{out}U_\mathrm{in})+P(0_\mathrm{M}U_\mathrm{out}L_\mathrm{in})+P(0_\mathrm{M}L_\mathrm{out}L_\mathrm{in})\big] \label{eq:setup_conditionalprobability_falsenegative}
\end{align}
As in the main text, we associate the flag result $0_\mathrm{M}$ with an \textit{unleaked} input that is transferred to an \textit{unleaked} output ($\mathrm{U}_\mathrm{out}\mathrm{U}_\mathrm{in}$) and result $1_\mathrm{M}$ with a \textit{leaked} input which is reset to an \textit{unleaked} output ($\mathrm{U}_\mathrm{out}\mathrm{L}_\mathrm{in}$), the probabilites of making the wrong guess given a flag result are.
Linking the probabilities of the flag measurement result $F$ to the gate transfer action $OI$ (mapping input type $I$ to output type $O$) is done via an intermediate measurement step, making use of ancillary singlet and triplet state being complementary events:
\begin{align}
&P(FOI) \nonumber\\ &= \sum_{j_\mathrm{A}\in\{S_\mathrm{A},T_\mathrm{A}\}}P(Fj_\mathrm{A}OI)=
\sum_{j_\mathrm{A}\in\{S_\mathrm{A},T_\mathrm{A}\}}P(F|j_\mathrm{A}OI)P(j_\mathrm{A}OI) \nonumber \\&\approx \sum_{j_\mathrm{A}\in\{S_\mathrm{A},T_\mathrm{A}\}}P(F|j_\mathrm{A})P(j_\mathrm{A}OI) \nonumber \\&= \sum_{j_\mathrm{A}\in\{S_\mathrm{A},T_\mathrm{A}\}}P(F|j_\mathrm{A})P(j_\mathrm{A}O|I)P(I)\label{eq:FlagInOut_fromconditionalprobabilities}
\end{align}
where we make the (reasonable) assumption that the measurement of the ancillary dots does only depend on the ancillary state $j_\mathrm{A}$ and not on the evolution ($OI$) due to the gate. We nominally identify the $F=0_\mathrm{M}$ flag result with a singlet $j_\mathrm{A}=S_\mathrm{A}$ and the $F=1_\mathrm{M}$ flag result with a singlet $j_\mathrm{A}=T_\mathrm{A}$, i.e. $P(1_\mathrm{M}|S_\mathrm{A})\equiv\epsilon_{1S}\ll 1$, $P(0_\mathrm{M}|S_\mathrm{A})=1-\epsilon_{1S}\approx 1$ and $P(0_\mathrm{M}|T_\mathrm{A})\equiv\epsilon_{0T}\ll 1$, $P(1_\mathrm{M}|T_\mathrm{A})=1-\epsilon_{0T}\approx 1$.

The relevant conditional probabilities of realizing a singlet or a triplet configuration in the ancillary spins paired with the computational dot state of interest are straightforwardly given by
\begin{align}
P(S_\mathrm{A}L_\mathrm{out}|U_\mathrm{in})&=0\\
P(T_\mathrm{A}L_\mathrm{out}|U_\mathrm{in})&=\bar{\epsilon}_\mathrm{L,ind}\\
P(S_\mathrm{A}U_\mathrm{out}|L_\mathrm{in})&=0\\
P(S_\mathrm{A}L_\mathrm{out}|L_\mathrm{in})&\equiv \epsilon_5 \ll 1\\
P(T_\mathrm{A}L_\mathrm{out}|L_\mathrm{in})&\equiv \epsilon_8 \ll 1\\
P(T_\mathrm{A}U_\mathrm{out}|L_\mathrm{in})&=1-\epsilon_5-\epsilon_8\approx 1
\end{align}
with $\epsilon_5 = \langle 5|\mathcal{E}(|5\rangle\langle 5|)|5\rangle$ and $\epsilon_8 = \langle 8|\mathcal{E}(|5\rangle\langle 5|)|8\rangle$ (see \textbf{Appendix \ref{sec:AppStates}}). This is only valid if errors arise exclusively from exchange noise, where all noisy gate actions are of Heisenberg type and have to conserve total angular momentum $J$. In particular, this implies $P(S_\mathrm{A}L_\mathrm{out}|U_\mathrm{in})=0$; this can be non-zero if noise processes that do not conserve total angular momentum are present.

Reintroducing the probability for pre-sequence leakage $P(L_\mathrm{in})\equiv\epsilon_\mathrm{L}\ll 1$ (and therefore $P(U_\mathrm{in})=1-\epsilon_\mathrm{L}$) from the main text and noting that the no-leakage flag result $F=0_\mathrm{M}$ should occur with high probability ($P(0_\mathrm{M})\approx 1$, implying $P(0_\mathrm{M}OI)\approx P(OI|0_\mathrm{M})$), inserting (\ref{eq:setup_conditionalprobability_falsenegative}) into \ref{eq:FlagInOut_fromconditionalprobabilities} straightforwardly yields
\begin{align}
P(L_\mathrm{out}U_\mathrm{in}|0_\mathrm{M})&\approx\epsilon_{0T}\bar{\epsilon}_\mathrm{L,ind}+P(S_\mathrm{A}L_\mathrm{out}|U_\mathrm{in})\\
P(U_\mathrm{out}L_\mathrm{in}|0_\mathrm{M})&\approx\epsilon_{0T}\epsilon_\mathrm{L}\\
P(L_\mathrm{out}L_\mathrm{in}|0_\mathrm{M})&\approx\epsilon_{5}\epsilon_\mathrm{L},
\end{align}
summing up to
\begin{align}
P(\overline{U_\mathrm{out}U_\mathrm{in}}|0_\mathrm{M})\approx \epsilon_{0T}\bar{\epsilon}_\mathrm{L,ind}+\epsilon_{0T}\epsilon_\mathrm{L}+\epsilon_{5}\epsilon_\mathrm{L}
\end{align}
to leading (second) order in the small error quantities due to (in order) inducing leakage and misidentifying a triplet as a singlet during measurement, having a leaked input and misidentifying a triplet as a singlet during measurement and having a leaked input and failing to remove leakage. Note the absence of the term due to inducing leakage and correctly measuring a singlet $P(0_\mathrm{M}|S_\mathrm{A})P(S_\mathrm{A}L_\mathrm{out}|U_\mathrm{in})P(U_\mathrm{in})\approx P(S_\mathrm{A}L_\mathrm{out}|U_\mathrm{in})$, which would have been of first order, because of the mentioned conservation of total angular momentum under strict Heisenberg noise.

The detection of incoming leakage will always occur with high reliability, essentially due to the low occurence of leakage in the first place. Same holds for leakage induction under Heisenberg noise, but may in principle be violated for arbitrary noise (e.g. magnetic gradient noise).

For the classification of false positives, the treatment is analogous. Starting with
\begin{align}
&P(\overline{U_\mathrm{out}L_\mathrm{in}}|1_\mathrm{M})\nonumber \\&= P(U_\mathrm{out}U_\mathrm{in}|1_\mathrm{M})+P(L_\mathrm{out}U_\mathrm{in}|1_\mathrm{M})+P(L_\mathrm{out}L_\mathrm{in}|1_\mathrm{M})
\nonumber\\
&=\frac{1}{P(1_\mathrm{M})}\big[P(1_\mathrm{M}U_\mathrm{out}U_\mathrm{in})+P(1_\mathrm{M}L_\mathrm{out}U_\mathrm{in})+P(1_\mathrm{M}L_\mathrm{out}L_\mathrm{in})\big] \label{eq:setup_conditionalprobability_falsepositive},
\end{align}
we specify the missing conditional probabilities via the error probabilities defined in the main text
\begin{align}
P(S_\mathrm{A}U_\mathrm{out}|U_\mathrm{in})&=1-\bar{\epsilon}_\mathrm{F}\\
P(T_\mathrm{A}U_\mathrm{out}|U_\mathrm{in})&=\bar{\epsilon}_\mathrm{F}-\bar{\epsilon}_\mathrm{L,ind}
\end{align}
Invoking (\ref{eq:FlagInOut_fromconditionalprobabilities}) again, one obtains
\begin{align}
P(1_\mathrm{M}U_\mathrm{out}U_\mathrm{in})&\approx\epsilon_{1S}+\bar{\epsilon}_\mathrm{F}-\bar{\epsilon}_\mathrm{L,ind}\\
P(1_\mathrm{M}L_\mathrm{out}U_\mathrm{in})&\approx\bar{\epsilon}_\mathrm{L,ind}\\
P(1_\mathrm{M}L_\mathrm{out}L_\mathrm{in})&\approx\epsilon_{8}\epsilon_\mathrm{L},
\end{align}
Contrary to the previous case, the measurement result $1_\mathrm{M}$ occurs rarely ($P(1_\mathrm{M})\ll 1$), more explicitly
\begin{align}
P(1_\mathrm{M}) &= P(1_\mathrm{M}U_\mathrm{out}U_\mathrm{in})+P(1_\mathrm{M}L_\mathrm{out}U_\mathrm{in})\nonumber \\&+P(1_\mathrm{M}L_\mathrm{out}L_\mathrm{in})+P(1_\mathrm{M}U_\mathrm{out}L_\mathrm{in})\nonumber\\
&\approx \epsilon_{1S}+\bar{\epsilon}_\mathrm{F}+\epsilon_\mathrm{L}
\end{align}
using that $P(1_\mathrm{M}U_\mathrm{out}L_\mathrm{in})\approx \epsilon_\mathrm{L}$ and dropping the second order contribution by $P(1_\mathrm{M}L_\mathrm{out}L_\mathrm{in})$.
With this, (\ref{eq:setup_conditionalprobability_falsepositive}) reads 
\begin{align}
P(\overline{U_\mathrm{out}L_\mathrm{in}}|1_\mathrm{M}) &\approx\frac{\epsilon_{1S}+\bar{\epsilon}_\mathrm{F}}{\epsilon_{1S}+\bar{\epsilon}_\mathrm{F}+\epsilon_\mathrm{L}} = \frac{1}{1+\epsilon_\mathrm{L}/(\epsilon_{1S}+\bar{\epsilon}_\mathrm{F})},
\end{align}
simply stating that if the error in flag extraction (either noise in the sequence or the measurement) $\epsilon_{1S}+\bar{\epsilon}_\mathrm{F}$ exceeds the average leakage probability $\epsilon_\mathrm{L}$, flag information becomes unreliable.

\section{Flagging and Quantum Error Correction}
\label{sec:AppFlagQEC}

In the following, we will consider the topic of leakage in the QEC context and offer a perspective on the situation specific to EO qubits.

The envisioned setting for QEC involves a separation of roles of physical qubits into data qubits, which collectively host the encoded quantum information, and ancillary qubits, employed to realize the stabilizer measurements necessary to detect errors on the data qubits \cite{Terhal2015}. The presence of leaked qubits will degrade the performance of such error correcting codes \cite{Fowler2013}, and as leakage will tend to accumulate under continued operation of such a system, leakage has to be actively removed.
\begin{figure}%[ht!]
	%\centering
    \includegraphics[scale=0.25]{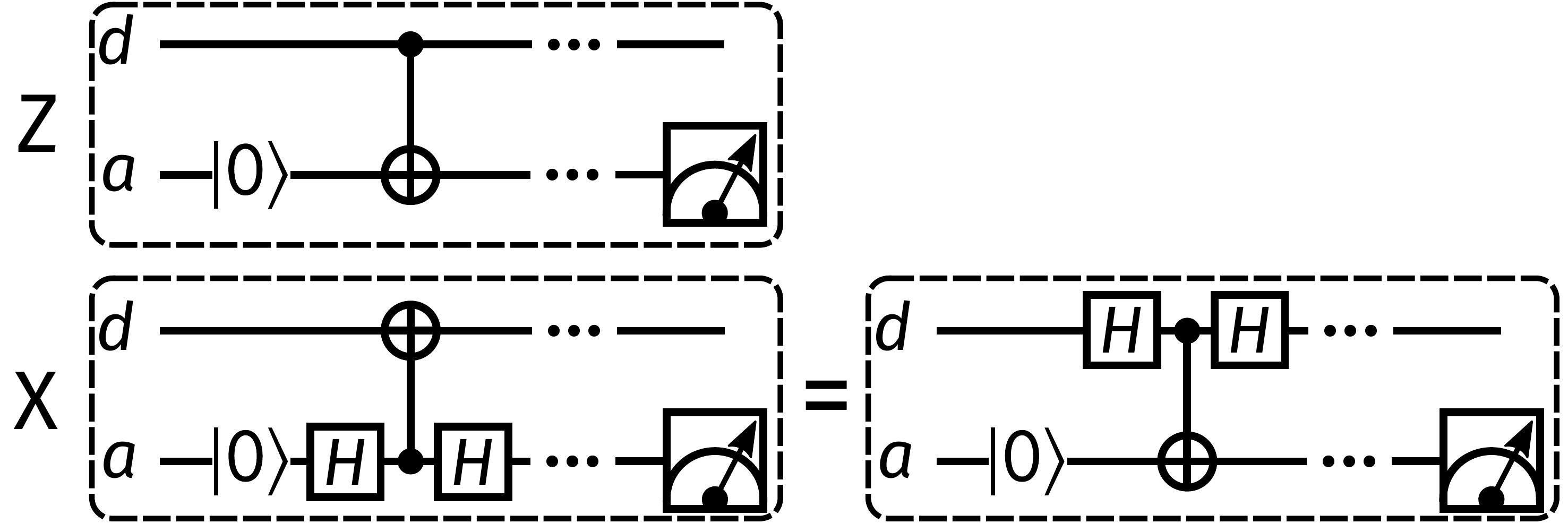}
	\caption{Schematic illustrating the actions executed between data and ancilla qubits during a single $Z$- (top) and $X$-stabilizer (bottom) round, probing the data qubit for an $X$-error and $Z$-error respectively. Dots ($\cdots$) serve as a placeholder for equivalent interactions carried out between different pairs of data and ancilla qubits until the stabilizer interaction is completed. For $X$-type stabilizers, two circuits are depicted which are equivalent on the qubit level, but may act differently when leakage is involved, depending on the two qubit gate employed.}
	\label{fig:AppStabilizerInteraction}
\end{figure}
Depicted in \textbf{Fig. \ref{fig:AppStabilizerInteraction}} are the operations involving data and ancilla qubits required for a single syndrome aquisition cycle. In particular, such a cycle involves ancilla initalization, maximally entangling two qubit interactions between ancilla and data qubits and ancilla qubit readout.

At the beginning of each stabilizer round, the ancilla qubit should be reset into a known initial state (e.g. computational $|0\rangle$). If this qubit reset also removes leakage, ancillary qubits will suffer from leakage for at most one code cycle and persistent leakage will only be present in data qubits. Under this condition, frequent interchange of locations of data and ancilla qubits allows for drainage of leakage from all physical qubits \cite{Suchara2015,Ghosh2015}.
However, the gates required for this have to be either SWAP-if-leaked or SWAP-if-not-leaked gates to leave the qubit information intact while resetting qubit leakage. For EO qubits, SWAP-if-leaked gates have been found to be twice as complex as the RiL procedure presented in this paper  \cite{FongWandzura2011} and similar complexity can be expected for a SWAP-if-not-leaked gate, as it may be generated by addition of a relatively short sequence of spin swaps from the former.

When qubit locations remain fixed, there is still merit in restricting the propagation of leakage between neighbouring qubits: Even if ancillary qubits are reset after each error correcting cycle, leakage could propagate from one data qubit to another during one cycle via interaction with a shared ancillary qubit or the leaked ancilla qubit may simply cause qubit faults via interaction with other data qubits before it is reset. A suitable choice of physical two qubit gates and their arrangement allows for suppression of these effects \cite{Brown2020}, which we shall now discuss for the EO setting.

In the EO case, exchange gate constructions allow, in principle, for universal control in all $J$ subspaces for an EO two qubit gate, so in principle an arbitrary interaction under leaked inputs can be straightforwardly engineered. However, it is worth pointing out that the short CNOT sequence found by Fong and Wandzura \cite{FongWandzura2011} is already a suitable candidate for such two qubit interactions: This CNOT sequence performs an $X$-gate on the target if the control is in a computational $|1\rangle$ \textit{or} a leaked state. On the other hand, a leaked target subjected to such a CNOT would entangle the qubit states with the gauge, as the evolution in the $J=1$ and $J=2$ subspaces is different, causing the evolution to be stochastic if one assumes the gauge states to be mixed. The stochastic action on the the control subjects it to erroneous gates or transfer into a leaked state.
Under these considerations, the circuit arrangement proposed in \cite{Brown2020}, motivated by the similar asymmetry in action w.r.t. leaked inputs of (cross-resonance) transmon CNOTs, may therefore be readily employed for the EO setting as well: Data qubits are designated as the control qubit for all two qubit interactions, corresponding to substitution of the circuit usually depicted for $X$-stabilizer interactions in \textbf{Fig. \ref{fig:AppStabilizerInteraction}} (bottom-left) by the one in \textbf{Fig. \ref{fig:AppStabilizerInteraction}} (bottom-right) and choosing the Fong-Wandzura construction as the CNOT. In this configuration, leaked data qubits never spread their leakage onto ancilla qubits and only perform erroneous $X$-gates on them, which also do not propagate from the ancilla qubit to data qubits involved in future interactions if the ancilla is always chosen as the target of a CNOT.

With the assertion that any EO single qubit gate acts trivially on a leaked state, due to total spin angular momentum conservation of the Heisenberg interaction, this circuit arrangement leads to qubit defect-like behavior of a leaked qubit \cite{Varbanov2020,Auger2017}, and may be unmasked by decoding stabilizer measurements as studied in \cite{Varbanov2020}. This may already be a powerful tool in conjunction with non-flagging RiL procedures, as expected presence of leakage before application of the unit and absence of leakage after the application may be compared with the syndrome history. The flag information would increase confidence in or may be substituted for the syndrome based inference; to what extent this applies and if it provides a significant boost to logical error mitigation remains to be clarified. The possibility of inferring leakage from stabilizer information may also open the route to targeted application of LRUs to qubits which are suspected to be leaked, but this would require a negligible delay in decoding and feedback times.

Finally, if ancilla reset is not performed in a manner that eliminates leakage (e.g. performing reset by singlet-triplet measurement), a persisting triplet measurement result may be probed for leakage (e.g. following a triplet measurement result by an X-gate and another measurement discriminates between computational 1 (singlet) or leakage state (triplet) via the second measurement result \cite{Andrews2019}), or directly subjected to an LRU out of suspicion.
As the excess total angular momentum has to be removed at some point, at least the ancillary spins used in the LRU require the capability of being reliably reset into a singlet state. 
This can occur by interaction with a fermion (electron) reservoir or by any other fiducial reset operation, erasing all information in the ancillary spins.

In short, one may cope with leakage in a QEC setting by choosing a more benign arrangement of target and control for stabilizer two-qubit gates in such codes, and on a higher level decoders may be optimized to deal with leakage, where flag information may prove useful as shown in \cite{Suchara2015}. This may be studied further by taking into account reliability of flag information as in Eqs. (\ref{eq:FalsePositive}-\ref{eq:FalseNegative}) as well as models for non-leakage errors and realistic singlet-triplet measurements \cite{Rispler2020}, all of which is beyond the scope of this paper.

\end{document}